\documentclass[journal]{IEEEtran}
\usepackage{cite}

\ifCLASSINFOpdf
  \usepackage[pdftex]{graphicx}
  \graphicspath{{Figures/}}
\else
\fi
\usepackage{amsmath}
\ifCLASSOPTIONcompsoc
  \usepackage[caption=false,font=normalsize,labelfont=sf,textfont=sf]{subfig}
\else
  \usepackage[caption=false,font=footnotesize]{subfig}
\fi
\usepackage{url}

\begin{document}
%
\title{
Huffman-Coded Sphere Shaping for \\Extended-Reach Single-Span Links
}

%
%
%

\author{Pavel~Skvortcov,~\IEEEmembership{Student~Member,~IEEE,}
        Ian~Phillips,~\IEEEmembership{Member,~IEEE,}
        Wladek~Forysiak,~\IEEEmembership{Member,~IEEE,}
        Toshiaki~Koike-Akino,~\IEEEmembership{Senior~Member,~IEEE,}
        Keisuke~Kojima,~\IEEEmembership{Senior~Member,~IEEE,}
        Kieran~Parsons,~\IEEEmembership{Senior~Member,~IEEE,} and~David~S.~Millar,~\IEEEmembership{Member,~IEEE}
\thanks{P. Skvortcov, I. Phillips, W. Forysiak are with Aston University, Birmingham, B4 7ET, UK. E-mails: \{skvortcp, i.phillips, w.forysiak\}@aston.ac.uk.}
\thanks{D. S. Millar, T. Koike-Akino, K. Kojima and K. Parsons are with Mitsubishi
Electric Research Laboratories (MERL), Cambridge, MA 02139, USA. E-mails: \{millar, koike, kojima, parsons\}@merl.com.}
}
\maketitle

\begin{abstract}
Huffman-coded sphere shaping (HCSS) is an algorithm for finite-length probabilistic constellation shaping, which provides nearly optimal energy efficiency at low implementation complexity. In this paper, we experimentally study the nonlinear performance of HCSS employing dual-polarization 64-ary quadrature amplitude modulation (DP-64QAM) in an extended-reach single-span link comprising 200~km of standard single-mode fiber (SSMF). We investigate the effects of shaping sequence length, dimensionality of symbol mapping, and shaping rate. We determine that the na\"{i}ve approach of Maxwell--Boltzmann distribution matching --- which is optimal in the additive white Gaussian noise channel --- provides a maximum achievable information rate gain of 0.18~bits/4D-symbol in the infinite length regime. Conversely, HCSS can achieve a gain of 0.37~bits/4D-symbol using amplitude sequence lengths of 32, which may be implemented without multiplications, using integer comparison and addition operations only. Coded system performance, with a net data rate of approximately 425~Gb/s for both shaped and uniform inputs, is also analyzed. 

\end{abstract}

\begin{IEEEkeywords}
Optical fiber communication, probabilistic shaping, sphere shaping, nonlinear fiber channel, single-span links.
\end{IEEEkeywords}

%
\IEEEpeerreviewmaketitle

\section{Introduction}
%
%
%
%
\bstctlcite{IEEEexample:BSTcontrol}
\IEEEPARstart{I}{n} the last decade, coherent detection has been a key enabling technology for high-throughput optical fiber communication systems. Initially, the adoption of coherent detection with high-speed digital signal processing (DSP) provided immediate four-fold improvement in spectral efficiency compared to direct detection systems, as all four dimensions of the optical field could be detected. The first coherent 40~Gb/s transmission systems using dual polarization (DP) quadrature phase-shift keying (QPSK) \cite{HanSun_40G} were therefore able to operate without significant increase in bandwidth compared with 10~Gb/s on-off keying systems. DP-QPSK 100~Gb/s systems were subsequently introduced \cite{100G_Demo}, and are widely deployed in commercial long-haul networks. In the presence of increasingly stringent bandwidth constraints, Nyquist pulse shaping and $M$-ary quadrature amplitude modulation ($M$QAM) have been employed in the next generations of coherent systems to increase spectral efficiency, and therefore achieve the required growth in per-wavelength bit rates \cite{QAM_FieldTrial_BTCiena2014}. 

Due to the reduction in noise tolerance of high spectral efficiency transmission systems, coded modulation (CM) schemes have been proposed to maximize the transmission system performance given the constraints of the physical channel.



By considering the optical field as a 4D signal space, power efficiency may be improved in a number of ways (e.g., by considering constellation points on an optimal lattice bound by a maximum power) \cite{4D_Agrell,4D_Alvorado}. High-dimensional modulation (HDM) based on utilization of multiple time-slots (consequent transmitted symbols) or multiple carriers to utilize more efficient lattices in larger numbers of dimensions can provide further improvement in power efficiency \cite{Millar_Modulation,Millar_Modulation_Invited}. However, coded performance using binary forward error correction (FEC) codes and bit-interleaved coded modulation (BICM) implementation for such systems may be challenging due to their lack of Gray-coded bit labeling. Additionally, while much research in this area has focused on power efficiency, several proposed HDM techniques specifically target improved nonlinear tolerance in the optical fiber channel \cite{Kojima_4D_NLin,Ciena_8D,Kojima_4D_SubCarrier}.   


Constellation shaping may be considered as the optimization of transmitted symbols distribution in terms of either location or probability in the signal space, such that the transmitted signal has improved power efficiency or nonlinear tolerance for an optical channel. In the additive white Gaussian noise (AWGN) channel, constellation shaping provides a  gain of up to $1.53$~dB in power efficiency over uniform signaling \cite{Kschischang_Shaping}. 

Finite-length probabilistic shaping may be considered as HDM --- i.e., the mapping of a block of $k$ input bits to a point on an $L$-dimensional constellation (e.g., on a square lattice), where an $L$-dimensional constellation point is then mapped to a sequence of 4D symbols in time (e.g., DP-$M$QAM) \cite{Laroia_ShapingMultiDim}. 

The key enabler for probabilistic shaping in practical systems was the introduction of a probabilistic amplitude shaping (PAS) framework for QAM \cite{Boecherer_PAS}. The PAS structure calls for blocks of uncoded input bits to be mapped onto blocks of amplitudes with some desirable probability distribution. The bits corresponding to the labels of the shaped amplitudes are then used as the input to a systematic FEC code. The parity bits (which are uniformly distributed) generated by the FEC code are then used to form the sign bits of the constellation, such that the resulting encoded distribution is symmetric about zero. Some information bits may be carried on the sign of the constellation if required, but this is not necessary. We note that one useful feature of such systems is that the overall transmission rate may be tuned by adjusting the rate of the shaping algorithms, while the FEC code rate remains the same.



Since the proposal of PAS, many implementations utilizing this structure have been investigated. Initially, an analysis based on an asymptotic infinite-length shaping approach with an ideal Maxwell--Boltzmann (MB) distribution of amplitudes was commonly performed for transmission over a nonlinear optical channel. MB shaping achieves the maximum power efficiency for a given entropy, which is therefore the optimal distribution in the AWGN channel \cite{Fischer_Book}. However, for an optical channel, MB shaping can enhance fiber Kerr nonlinearities, leading to a decrease in shaping gain, which was shown for long-haul transmission \cite{Fehenberger_JLT06} and unrepeated transmission \cite{Renner_JLT}. Importantly, while considerable performance improvement can be achieved using infinite-length MB shaping, this approach is fundamentally impossible to implement. 

More recently, research has focused on finite-length shaping architectures which can be realizable in the hardware. One class, referred to as distribution matching (DM), is based on obtaining a fixed target distribution. First, constant-composition DM (CCDM) was introduced \cite{CCDM}, whereby all transmitted amplitude sequences are permutations of a single composition, defined by the target distribution. While CCDM can provide asymptotically low rate-loss, and therefore high power efficiency, it requires long sequences (typically on the order of several hundred amplitudes) to achieve it. Multiset-partition DM (MPDM) was proposed in \cite{Fehenberger_MPDM_TCom,Millar_MPDM} and provides lower rate loss at a fixed shaping sequence length. MPDM is based on multiple complimentary compositions, which on average result in the desired distribution. Other DM implementations were also proposed, such as product DM \cite{Product_DM}, hierarchical DM \cite{Hierarchical_DM}, prefix-free code DM \cite{CCDM}, and parallel-amplitude DM \cite{Parallel_DM}. 

Another class of shaping architectures is based on sphere shaping. Sphere shaping introduces an optimal sphere bound in multi-dimensional signal space --- for a target transmission rate and finite sequence length, the optimal set of constellation points from a fixed lattice is chosen such that the geometry is bounded by a hyper-sphere \cite{Laroia_ShapingMultiDim,Fischer_Book}. This scheme has intuitively optimal power efficiency --- for any rate, lattice and sequence length, we can define a smallest possible sphere which contains the correct number of points: any alternative set of constellation points will contain points outside the sphere, leading to degraded power efficiency. Sphere shaping-based algorithms include shell mapping (SM) \cite{ShellMapping, ShellMapping2}, enumerative sphere shaping (ESS) \cite{ESS,ESS2,ESS3}, and Huffman-coded sphere shaping (HCSS) \cite{Millar_ECOC19,Fehenberger_JLT20}.    

The dependence of nonlinear tolerance on shaping length was investigated using constant composition distribution matching (CCDM) \cite{Fehenberger_OFC20,Fehenberger_ShortLength_CCDM} and ESS \cite{ESS_Exp,ESS3} for multi-span long-haul links. Also, the advantage of 2D symbol mapping for nonlinearity tolerance has been mentioned previously \cite{Fehenberger_ShortLength_CCDM}, while the advantage of the short-length shaping regime for a nonlinear optical channel was theoretically investigated in \cite{Dar_OnShapingNLin,Geller_JLT}. Significant shaping gain exceeding the theoretical gain for the AWGN channel was demonstrated in numerical simulations for single-span links by optimally combining linear and nonlinear shaping gains using SM with very short shaping lengths \cite{Geller_JLT}. In \cite{Mapping_4D,Mapping_4D_Exp} the authors investigated shaping of a single 4D quadrant using SM and demonstrated increased nonlinear tolerance in single-span links. In our previous work, we demonstrated a significant nonlinear shaping gain with short-length HCSS and 4D amplitude-to-symbol mapping for extended-reach single-span links \cite{HCSS_Exp}.

Nonlinearity tolerant shaping is of particular interest for short-distance transmission with low accumulated chromatic dispersion (CD). Low CD leads to highly correlated short-memory nonlinear interactions. Hence, for short-distance systems with improved nonlinear tolerance can be achieved with short-length shaping \cite{Geller_JLT}, which is attractive in terms of implementation complexity. In contrast, for long-haul systems nonlinearities become significantly decorrelated (turning into Gaussian-like noise) and longer length shaping is required, as for AWGN channels.   

At present, demand for high capacity single-span transmission systems is driven by cloud and inter-data-center traffic. After successful introduction of the 400G ZR standard \cite{OIF_400ZR}, which specifies transmission of up to $120$~km, there is great interest in increasing system reach and bit rate. When increasing the transmission distance of the single-span system, the signal becomes significantly impaired by fiber nonlinearities due to increased optimal launch power. As demonstrated previously, for these systems short-length shaping can offer significant performance gain at reasonably low implementation complexity \cite{HCSS_Exp}. 

In this work we extend our study on nonlinear performance of HCSS in extended-reach single-span links. We investigate the impact of shaping length, dimensionality of symbol mapping, and rate of HCSS on nonlinear tolerance in comparison with uniform signaling and infinite-length shaping with an ideal MB distribution. Also, coded performance is analyzed using low-density parity-check (LDPC) codes, which provide matching data rate for shaping schemes and uniform signaling.

The remainder of this paper is structured as follows. In Section \ref{sec:hcss} we give an overview of the PAS framework, HCSS architecture, multi-dimensional symbol mapping strategies, associated energy efficiency and rate loss of shaping/mapping scheme. Section \ref{sec:exp} provides the description of the transmission experiment, DSP and performance metrics used for evaluating system performance and comparing shaping methods. In Section \ref{sec:result} we present the experimental results and discuss the nonlinear system performance, while in Section \ref{sec:conc} we draw our conclusions. Appendices \ref{FirstAppendix} and \ref{SecondAppendix} give detailed explanations of probability mass function (PMF) calculation in  high-dimensional symbol mapping, and experimental data fitting used for the analysis of the results, respectively.

\section{Probabilistic Shaping: Huffman-Coded Sphere Shaping (HCSS)}
\label{sec:hcss}

\begin{figure}[!t]
\centering
\captionsetup[subfloat]{farskip=1pt}
\subfloat[Transmitter]{\includegraphics[width=1\linewidth]{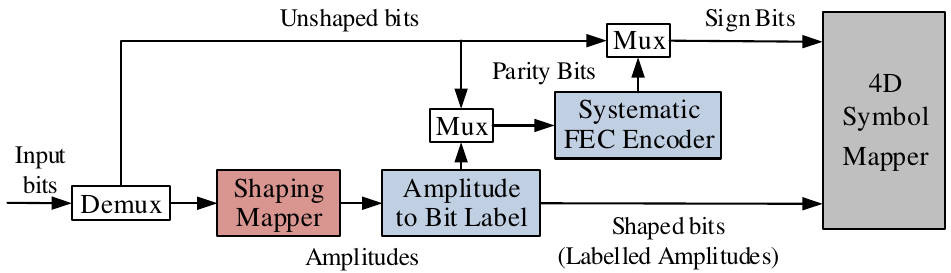}}
\hfil
\subfloat[Receiver]{\includegraphics[width=1\linewidth]{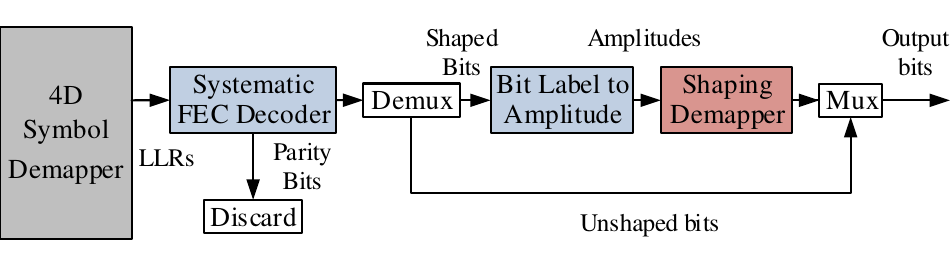}}
\caption{PAS architecture at the transmitter and receiver.}
\label{fig:PAS}
\end{figure}

\subsection{Probabilistic amplitude shaping (PAS)}

PAS is a shaping technique whereby blocks of information bits are mapped onto probabilistically shaped amplitude sequences\cite{Boecherer_PAS}. The bit labels corresponding to the shaped amplitudes are then encoded with an FEC code, and the parity bits assigned (in some cases, along with some information bits) to the sign bits of the pulse amplitude modulation (PAM) constellation. These signed amplitudes are then mapped onto the 4D optical carrier for transmission. The primary advantage of the PAS architecture is that the FEC decoder operates on the bit labels of the shaped sequences at the receiver, enabling FEC decoding to be performed before shaping demapping. The demapping is then performed on amplitudes which are presumed to be error free, greatly reducing the complexity of both the demapping procedure and the shaping system design.

The diagram of PAS architecture at the transmitter and receiver is shown in Fig.~\ref{fig:PAS}.
We define the shaping rate in bits per amplitude (b/Amp) as 
\begin{equation}
\label{eq:shapingrate}
R_{\mathrm{S}} = \frac{k}{L}\,,
\end{equation}
where $k$ is the number of uniform input bits, and $L$ is the length of the shaped amplitude sequence. 

\subsection{Huffman-coded sphere shaping (HCSS)}

\begin{figure}[!t]
\centering
\captionsetup[subfloat]{farskip=1pt}
\subfloat[HCSS Mapper]{\includegraphics[scale = 0.34]{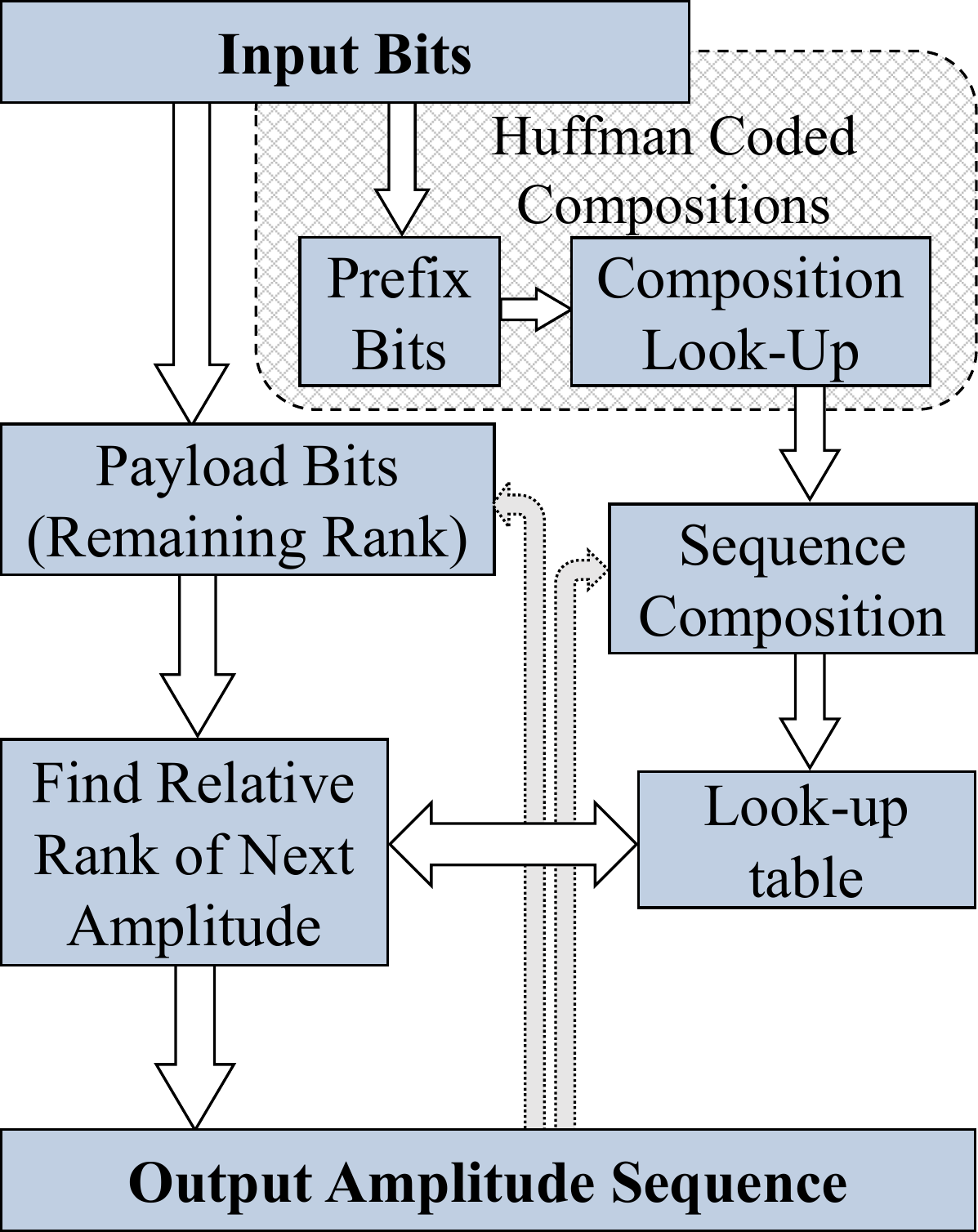}}
\hfil
\subfloat[HCSS Demapper]{\includegraphics[scale = 0.34]{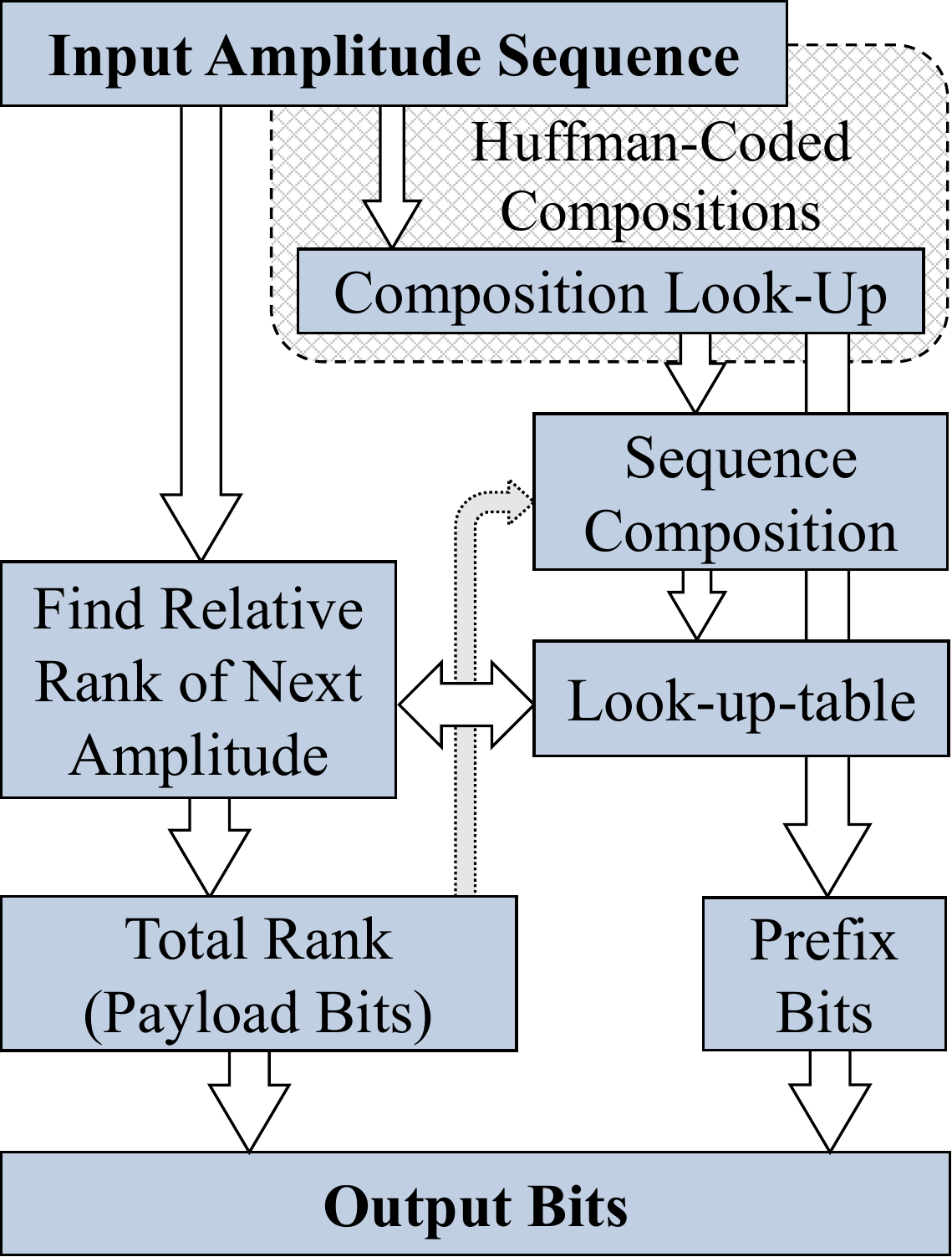}}
\caption{Block diagram of HCSS algorithm using multiset ranking and LUT.}
\label{fig:HCSS_Block}
\end{figure}

The sphere bound on power efficiency for fixed-length shaping utilizes all constellation points for signaling on a specified multi-dimensional lattice having a certain energy constraint. By definition, this scheme achieves the best possible energy efficiency for a given rate (i.e., number of constellation points), lattice (e.g., the square lattice modulation of uniform QAM), and dimension (i.e., sequence length)\cite{Fischer_Book}. HCSS restricts the number of constellation points utilized for each unique composition to be a power of two, and then introduces a minimal number of additional compositions with higher power to ensure a dyadic distribution of compositions \cite{Millar_ECOC19}. This enables the use of a variable length binary prefix (Huffman code) to uniquely address compositions in the shaping architecture. The remaining payload bits in a binary input word are then used to address a unique permutation of the specified composition. If the multiset ranking (MR) algorithm is used as described in \cite{Millar_ECOC19,Fehenberger_JLT20}, the lexicographical rank of the selected sequence corresponds to the payload bits. Additionally, we note that sequence ranks may be computed without multiplication operations by pre-computing multinomial coefficients and storing them in a look-up table (LUT). For short sequence lengths, the coefficients required for MR mapping and demapping can be stored in moderately sized LUTs --- for example, sequence length of $32$ requires only $100$~kbits of memory\cite{Fehenberger_JLT20}. Therefore, both mapping in Fig.~\ref{fig:HCSS_Block}(a) and demapping in Fig.~\ref{fig:HCSS_Block}(b) are performed iteratively on a per-symbol basis, using LUTs, requiring integer comparison and addition operations only. 

The number of available sequences for the $i^\mathrm{th}$ composition in the Huffman-coded structure is
\begin{equation}
\label{eq:numbersequences}
N^i_{\mathrm{seq}} = 2^{\lfloor  \log_{2} N^i_{\mathrm{perm}} \rfloor}\,,
\end{equation}
where $N^i_{\mathrm{perm}}$ is the number of possible permutations of the $i^\mathrm{th}$ composition, which is given by multinomial coefficient: 



\begin{equation}
    N^i_{\mathrm{perm}} = \frac{L!}{\prod_a{c^i(a)!}}\,,
\end{equation}
where $c^i(a)$ is the number of occurrences in the $i^\mathrm{th}$ composition of the amplitude $a$ for $L = \sum_a{c^i(a)}$.
The probability of the $i^\mathrm{th}$ composition is therefore given by: 
\begin{equation}
\label{compprobability}
p_i = \frac{N^i_{\mathrm{seq}}}{2^k}\,.
\end{equation}

\begin{figure}[!ht]
\centering
\captionsetup[subfloat]{farskip=1pt}
\subfloat[1D symbol mapping]{\includegraphics[scale = 1.03]{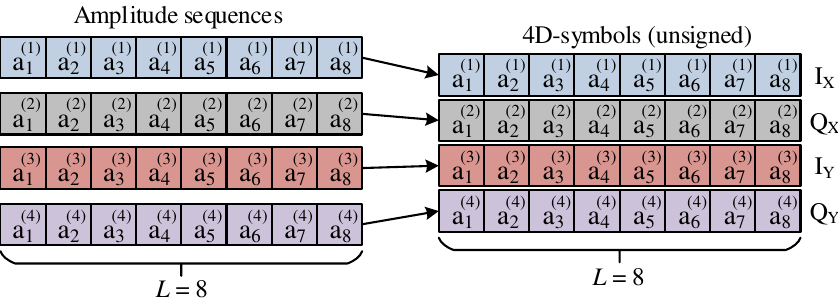}}\\
\subfloat[2D symbol mapping]{\includegraphics[scale = 1.03]{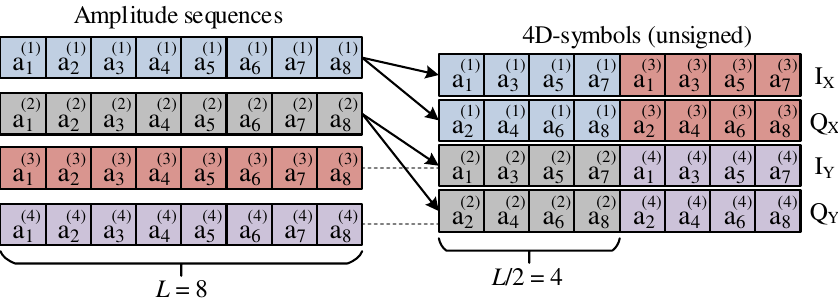}}\\
\subfloat[4D symbol mapping]{\includegraphics[scale = 1.03]{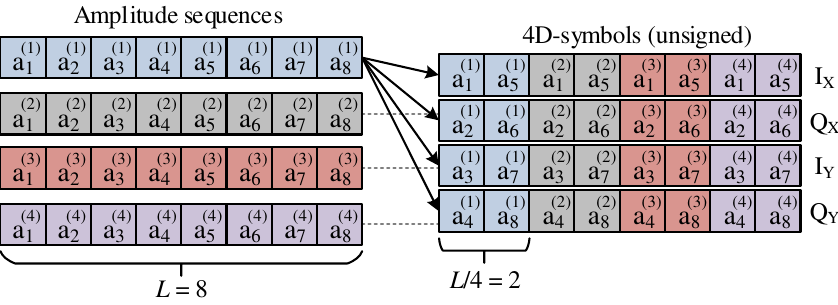}}
\caption{Strategies for mapping amplitude sequences into modulated 4D symbols (added signs are not reflected): (a) 1D symbol mapping, (b) 2D symbol mapping, (c) 4D symbol mapping.}
\label{fig:Mapping}
\end{figure}

\subsection{Symbol mapping strategies}
We studied three strategies for mapping shaped sequences of amplitudes into the modulated 4D symbols of the DP-64QAM format. These mapping strategies are illustrated in Fig.~\ref{fig:Mapping}. We note that a sign is assigned to each amplitude according to the sign bit during the mapping process (see Fig.~\ref{fig:PAS}~(a)), however, added signs are not reflected in Fig.~\ref{fig:Mapping} for simplicity of consideration.

We refer to 1D symbol mapping when four independent shaped amplitude sequences of length $L$ are sequentially (on an amplitude-by-amplitude basis) mapped into four simultaneous quadratures (in-phase and quadrature signal components in both polarizations) constructing a single 4D-symbol sequence of length $L$, as shown in Fig.~\ref{fig:Mapping}~(a). 

In the case of 2D symbol mapping, as shown in Fig.~\ref{fig:Mapping}~(b), two independent shaped amplitude sequences of length $L$ are mapped into a single 4D-symbol sequence of length $L/2$ --- two consecutive amplitudes from each amplitude sequence are mapped into four simultaneous quadratures.

In the case of 4D symbol mapping, as shown in Fig.~\ref{fig:Mapping}~(c), a single shaped amplitude sequence of length $L$ is mapped in a single 4D-symbol sequence of length $L/4$ --- four consecutive amplitudes are mapped into four simultaneous quadratures.

By increasing symbol mapping dimensionality we can effectively reduce the time-domain length of the output 4D-symbol sequence and the number of simultaneously interacting sequences, while maintaining the same power efficiency. The derivation of the resulting multi-dimensional PMF for different symbol mapping strategies is described in detail in Appendix~\ref{FirstAppendix}. We note that 1D, 2D and 4D mapping strategies result in 1D, 2D and 4D distributions, respectively.   

\subsection{Rate loss and power efficiency}
\begin{figure}[!t]
\centering
\captionsetup[subfloat]{farskip=1pt}
\includegraphics[scale = 0.50]{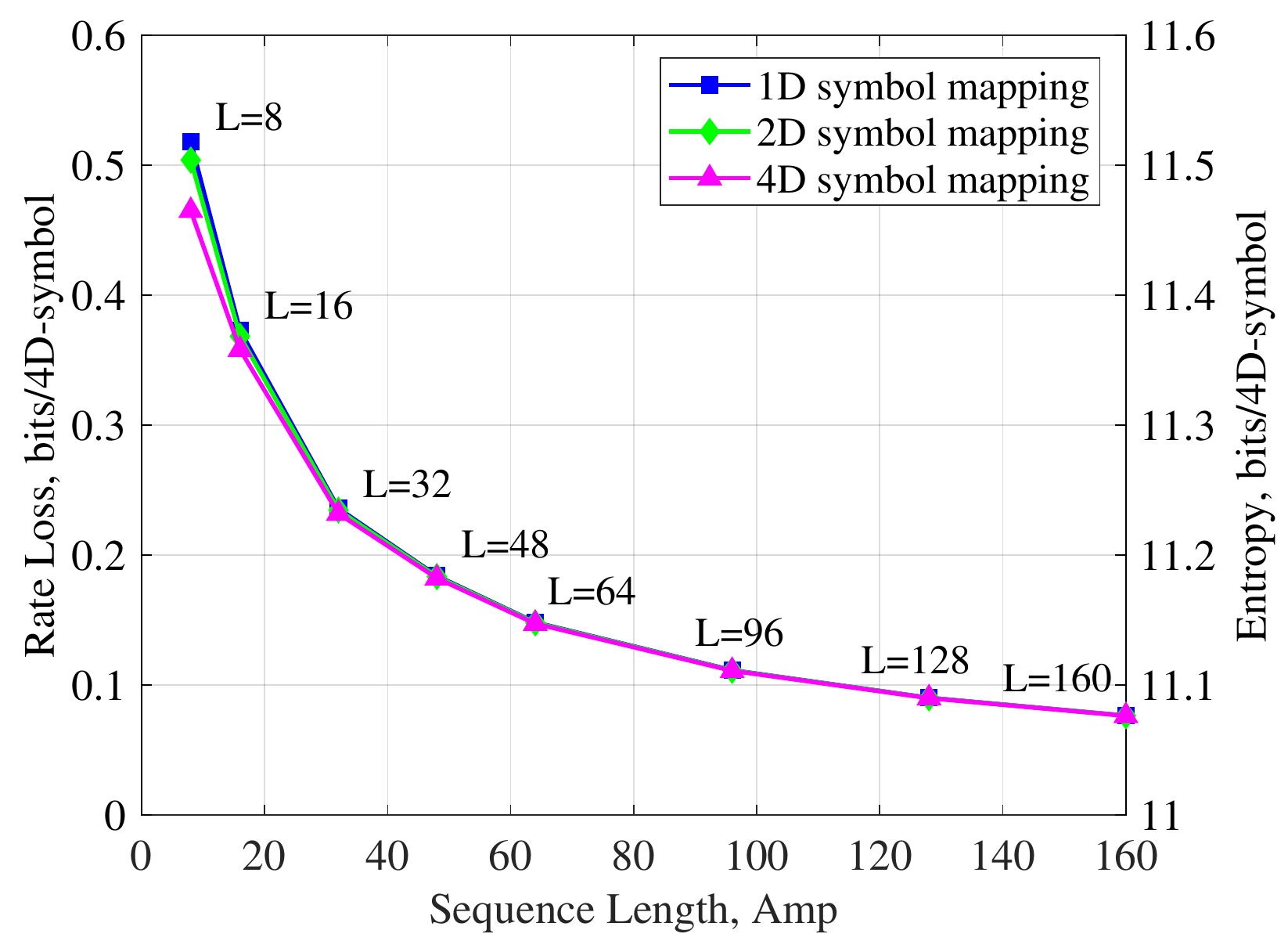}
\caption{Rate loss and entropy vs. shaping sequence length ($R_{\mathrm{S}} = 1.75$) for 1D, 2D and 4D symbol mapping strategies.}
\label{fig:RateLoss}
\end{figure}

Rate loss due to the use of a finite-length shaping scheme is calculated in bits per 4D-symbol (b/4D) as
\begin{equation}
\label{rateloss}
R_{\mathrm{loss}}=\mathrm{H}(X) - D\cdot(R_{\mathrm{S}}+1)\,,
\end{equation}
where $\mathrm{H}(X)$ denotes the entropy of the 4D output signal $X$ (calculated according to PMF in Appendix~\ref{FirstAppendix}), $D$ accounts for mapping dimensionality ($D=4$) and ``$1$'' accounts for the sign bit per dimension. Fig.~\ref{fig:RateLoss} demonstrates the rate loss and corresponding entropy using DP-64QAM in the case of 1D, 2D and 4D mapping at $R_{\mathrm{S}} = 1.75$. Higher dimensional mapping demonstrates slightly reduced rate loss at short shaping lengths.       

\begin{figure}[!t]
\centering
\captionsetup[subfloat]{farskip=1pt}
\includegraphics[scale = 0.50]{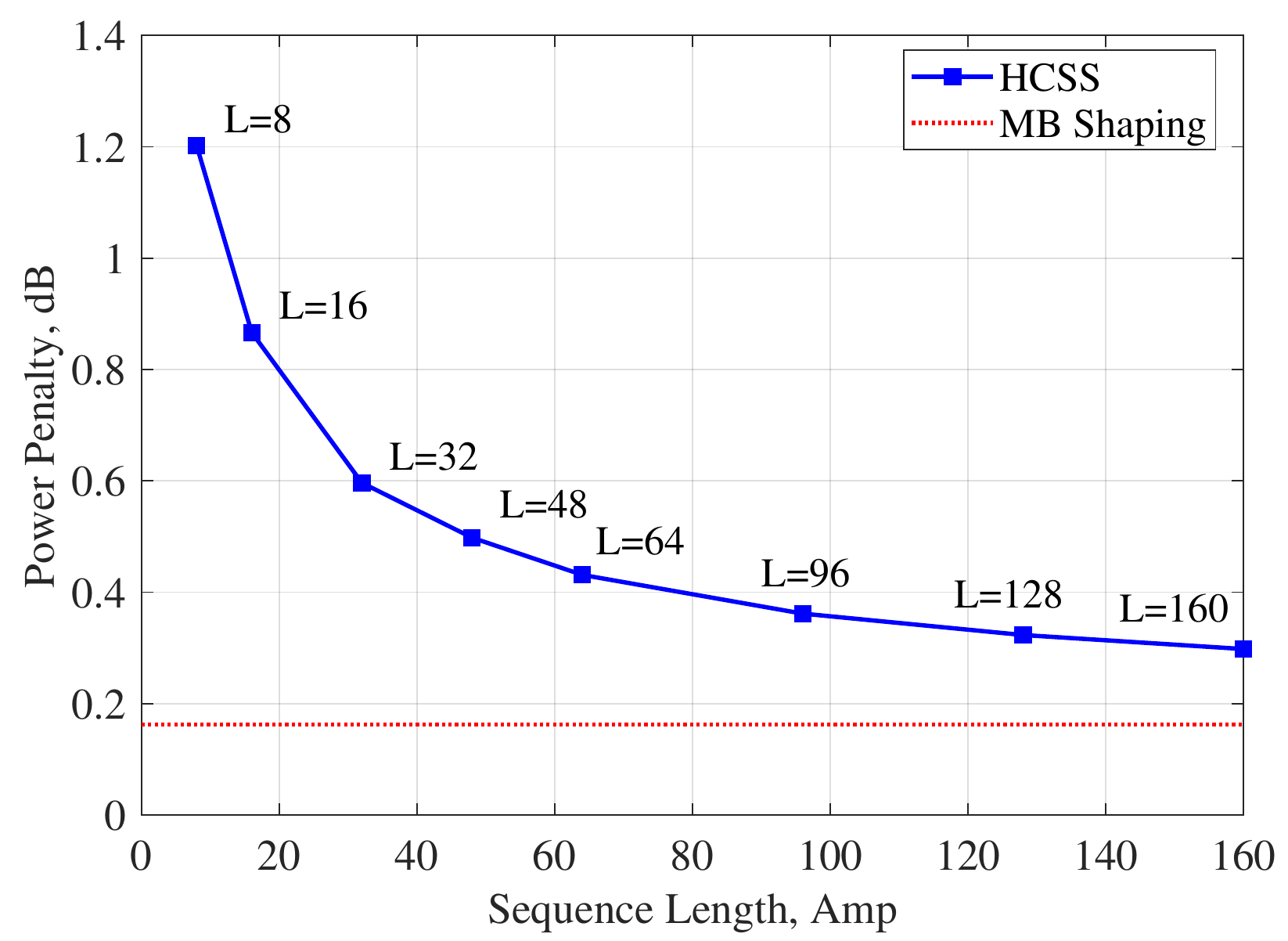}
\caption{Power penalty vs. shaping sequence length ($R_{\mathrm{S}} = 1.75$). Reference level is the MB shaping with unconstrained cardinality.}
\label{fig:PowerPenalty}
\end{figure}

Fig.~\ref{fig:PowerPenalty} shows the power penalty of HCSS and infinite-length MB shaping (also, using DP-64QAM) at $R_{\mathrm{S}} = 1.75$. Power penalty is calculated with respect to the MB distribution with unconstrained cardinality, which can be shown to provide the maximum power efficiency for a given entropy\cite{Fischer_Book}. We note that for HCSS power efficiency does not depend on symbol mapping dimensionality.

\section{Transmission Experiment}
\label{sec:exp}

We investigated the system performance using HCSS in comparison with uniform signaling and infinite-length MB shaping. For HCSS we varied the shaping sequence length ($L$ in the range of $8$--$160$), dimensionality of amplitude-to-symbol mapping (1D, 2D and 4D) and shaping rate ($R_{\mathrm{S}}$ in the range of $1.625$--$1.875$~b/Amp).



For the MB shaping case, signals were drawn from an MB distribution with the entropy matching the shaping rate of HCSS \cite{Fehenberger_JLT06}. We note that symbols in the transmitted signal were drawn independently and identically on the underlying PMF. This method may be considered to give a finite-length sample of an infinite-length shaped sequence, which incurs no rate loss.

\begin{figure*}[!t]
\centering
\includegraphics[width=0.98\textwidth]{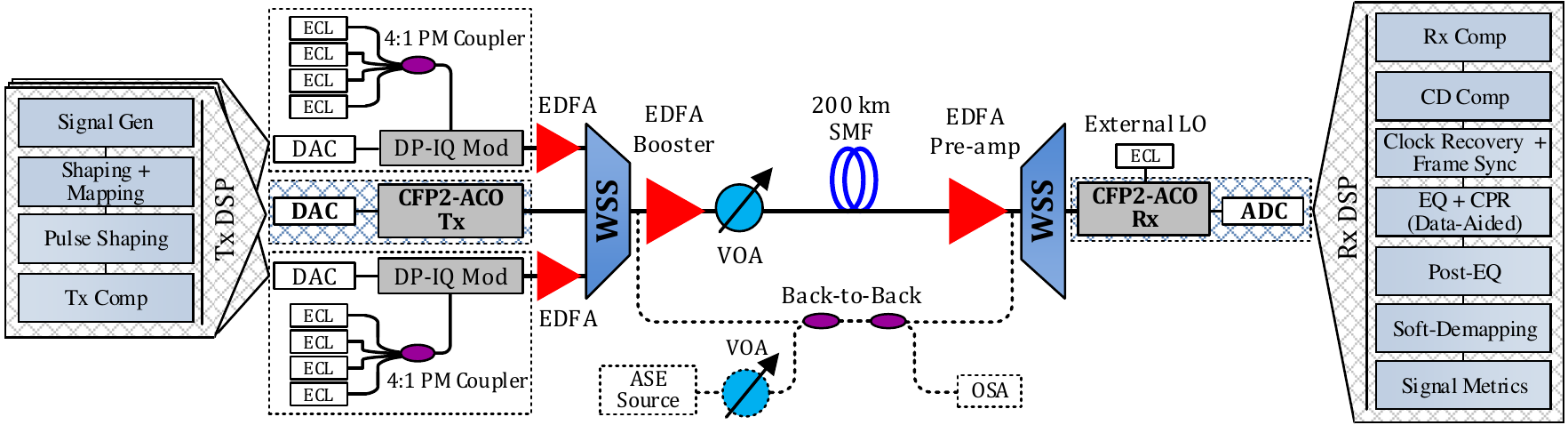}
\caption{Block diagram of transmission experimental setup. Block diagram of offline DSP for transmitter and receiver. 
}
\label{fig:mesh1}
\end{figure*}

\begin{figure}[!t]
\centering
\includegraphics[scale = 0.50]{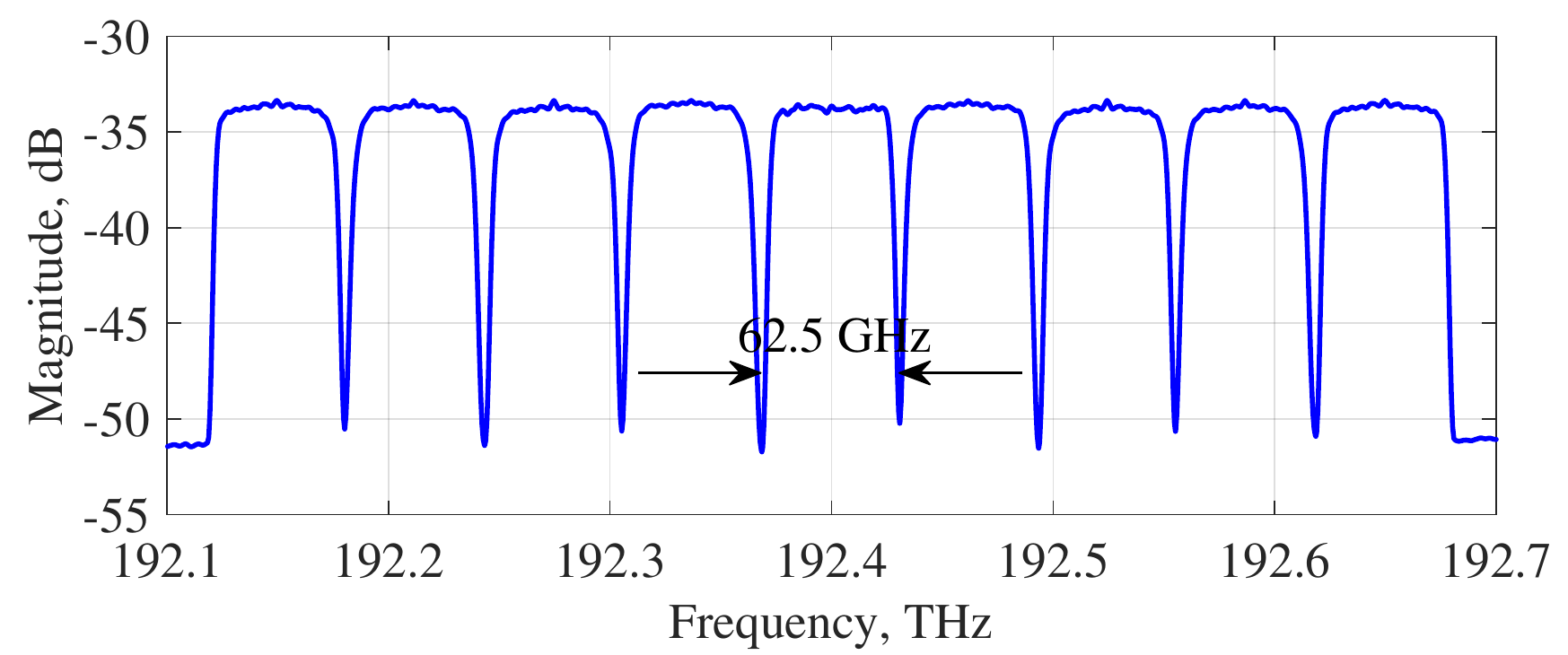}
\caption{Signal spectrum (measured with $2$~GHz resolution). $9$ channels (root-raised cosine pulse shaping with $10$\% roll-off factor) at $56$~GBd on a $62.5$~GHz grid.}
\label{fig:Spectrum}
\end{figure}

\subsection{Experimental setup}

The experimental setup is shown in Fig.~\ref{fig:mesh1}. Wavelength-division multiplexed (WDM) transmission of $9$ DP-64QAM channels operating at $56$~GBd (with root-raised cosine pulse shaping with $10$\% roll-off factor) on a $62.5$~GHz grid was carried out over a $200$~km single-span link of SSMF. All channels used the same shaping scheme under investigation. 

The central channel-under-test (CUT) was generated using a $92$~GSa/s digital-to-analog converter (DAC), followed by a pluggable CFP2 analog coherent optics (ACO) transceiver \cite{Forysiak_CFP2} (integrated laser with nominal $100$~kHz linewidth). The $8$ interfering channels were divided into two groups of $4$ channels, and generated by two pairs of DP Mach--Zehnder in-phase/quadrature-phase (IQ) modulators and DACs, where $4$ external cavity lasers with $100$~kHz linewidth combined with a polarization maintaining coupler were used for each modulator/DAC pair. The two groups of interfering channels were then pre-amplified by erbium-doped fiber amplifiers (EDFAs), and spectrally interleaved and combined with the CUT via a programmable wavelength selective switch (WSS). The resulting spectrum of the generated WDM signal is shown in Fig.~\ref{fig:Spectrum}.      

At the link input the WDM signal was amplified by a booster EDFA, then a variable optical attenuator (VOA) was used to control the total launched power. After transmission over $200$~km of SSMF, the WDM signal was pre-amplified by an EDFA with a noise figure of $5.5$~dB and the CUT was filtered by a WSS. Finally, the CUT was received by a CFP2-ACO transceiver (an external laser with $100$~kHz linewidth was used as a local oscillator) and its analog output was sampled and digitized by a $92$~GSa/s analog-to-digital converter (ADC). 

In back-to-back configuration, the output of the transmitter-side WSS was connected to the receiver-side WSS via a pair of couplers used for noise loading. An amplified spontaneous emission (ASE) noise source and VOA were used for varying optical signal-to-noise ratio (OSNR), while an optical spectrum analyzer (OSA) was used for OSNR measurement.

\subsection{Digital signal processing (DSP)}

DSP was performed offline according to the generic flow in Fig.~\ref{fig:mesh1}. At the transmitter-side, firstly, signals were randomly generated according to the shaping and mapping scheme under consideration. Then, root-raised cosine pulse shaping was applied followed by transmitter pre-emphasis (to compensate for frequency response and skews). Finally, the signals were uploaded to the DACs. 

At the receiver-side the signal recovery was implemented as follows. Firstly, the received data was extracted from the ADC and receiver pre-compensation was applied (compensation of frequency response, skews and I/Q imbalance). Then, CD was compensated, clock recovery was performed by a frequency-domain Gardner algorithm, conventional complex-valued decision-directed least-mean squares (DD-LMS) $2\times 2$ multiple-input multiple-output (MIMO) equalization ($35$ taps) was performed in conjunction with carrier phase recovery (CPR) in fully data-aided mode, and post-equalization was done using real-valued DD-LMS $2\times 2$ MIMO equalizers ($5$ taps) to enable compensation of residual transmitter IQ impairments \cite{Post_EQ}. Soft-demapping assumed a circularly symmetric Gaussian channel, and transmission performance metrics were averaged over approximately $5\times10^6$ symbols.

\subsection{Performance metrics}
We denote the transmitted signal by $X$ and received signal after DSP algorithms by $Y$. The transmitted signal $X$ takes values from the 4D-constellation $\mathcal{X} = \{x_1,\ldots,x_M\}$. We define the adjusted transmitted signal $X^{\prime}$ to take values from $\mathcal{X^{\prime}} = \{x_1^{\prime},\ldots,x_M^{\prime}\}$, such that the new constellation points $x_{i}^{\prime}$ represent the centroids of the received symbols $y$ corresponding to the original constellation points $x_i$, which can be expressed as      
\begin{equation}\label{centroid}
x^{\prime}_i = \mathrm{E}[y|x_i]\,,
\end{equation}
where $\mathrm{E}[\cdot]$ denotes the expectation. 

The adjusted constellation $\mathcal{X^{\prime}}$ is subsequently used for estimation of effective SNR and soft-demapping. This reduces the impact of impairments, which represent geometrical distortions of the constellation (e.g., transceiver nonlinearity and uncompensated modulation impairments). 

The effective SNR \cite{Fehenberger_JLT06} of the received signal is calculated as 
\begin{equation}
{\rm{SNR}_{eff}} = \frac{\mathrm{Var}[X^{\prime}]}{\mathrm{Var}[Y-X^{\prime}]}\,,
\label{snr}
\end{equation}
where $\mathrm{Var}[\cdot]$ denotes the variance. Effective SNR accounts for both linear and nonlinear noise contributions accumulated during signal propagation over an optical fiber, as well as transceiver noise floor (i.e., implementation penalty and DSP imperfections).

The soft-demapper calculates log-likelihood ratios (LLRs) for binary labels $B_i$ ($i=1,\ldots,m$) as 
\begin{equation}
\label{llr}
{\mathrm{LLR}}_{i} = \log_{}\frac{\sum_{x\in \mathcal{X}^{\prime}_{1,i} }f_{Y|X^{\prime}}(y|x^{\prime}) P_{X^{\prime}}(x^{\prime})}{\sum_{x\in \mathcal{X}^{\prime}_{0,i} }f_{Y|X^{\prime}}(y|x^{\prime}) P_{X^{\prime}}(x^{\prime})}\,,
\end{equation}
where $\mathcal{X}^{\prime}_{1,i}$ and $\mathcal{X}^{\prime}_{0,i}$ are the subsets of constellation $\mathcal{X}^{\prime}$, which represent $B_i$ being equal to $1$ or $0$, respectively; $f_{Y|X^{\prime}}(y|x^{\prime})$ is the transition probability density function of the auxiliary channel used for mismatched decoding; $P_{X^{\prime}}(x^{\prime})$ is the 4D PMF (calculated according to Appendix~\ref{FirstAppendix}).      

We consider a memoryless 4D circularly symmetric Gaussian auxiliary channel and assume that the noise in each dimension is independent and identically distributed \cite{Eriksson_4D}. In this case, the channel can be described as  
\begin{equation}
\label{channel}
f_{Y|X^{\prime}}(y|x^{\prime}) = \frac{1}{(\pi \sigma^2)^2} \exp\Big({-\frac{\|y-x^{\prime}\|^2}{\sigma^2}}\Big)\,,
\end{equation}
where $\sigma^2$ is the noise variance. We note that lower-dimensional soft-demapping can be done analogously for the case of 1D and 2D mapping.

The achievable information rate (AIR) for bit-metric decoding (BMD) impacted by the rate loss associated with finite-length shaping  \cite{Fehenberger_MPDM_TCom} is calculated in b/4D as 
\begin{equation}
\label{air}
{\mathrm{AIR}} =\underbrace{ \bigg [\mathrm{H}(X) - \sum_{i=1}^{m}\mathrm{H}(B_{i}|Y)  \bigg ]}_{R_{\mathrm{BMD}}} - R_{\mathrm{loss}}\,,
\end{equation}
where $R_{\mathrm{BMD}}$ is the BMD rate, which is given by the generalized mutual information (GMI). We note that for both uniform signaling and infinite-length MB shaping, we have $R_{\rm{loss}} = 0$.

Coded performance was analyzed based on normalized GMI ($\textrm{nGMI}$), which can be calculated for a uniform signal as 
\begin{equation}
\label{ngmi_uni}
{\mathrm{nGMI}} =\frac{\mathrm{GMI}}{m} = \frac{\mathrm{AIR}}{m}\,,
\end{equation}
while for shaped signals it is calculated as 
\begin{equation}
\label{ngmi_shaped}
{\mathrm{nGMI}} = 1 - \frac{\mathrm{H}(X) - \mathrm{GMI}}{m} = 1 - \frac{4\cdot(R_{\mathrm{S}}+1) - \mathrm{AIR}}{m}\,,
\end{equation}

For the shaped signals, an inner LDPC code with rate $0.72$ and length $52{,}800$ was used, and decoding was performed over $32$ iterations of the sum-product algorithm. Uniform signals were encoded with an LDPC code with rate $0.64$ and length $52{,}800$, and decoded in the same manner. In both cases, we assume the use of an outer Bose--Chaudhuri--Hocquenghem (BCH) code with rate $0.9922$, which achieves an output BER below $10^{-15}$ given an input BER of $5\times 10^{-5}$ \cite{Millar_1TbJLT}. 

\section{Experimental Results}
\label{sec:result}

\begin{figure}[!t]
\centering
\captionsetup[subfloat]{farskip=1pt}
\subfloat[]{\includegraphics[scale = 0.50]{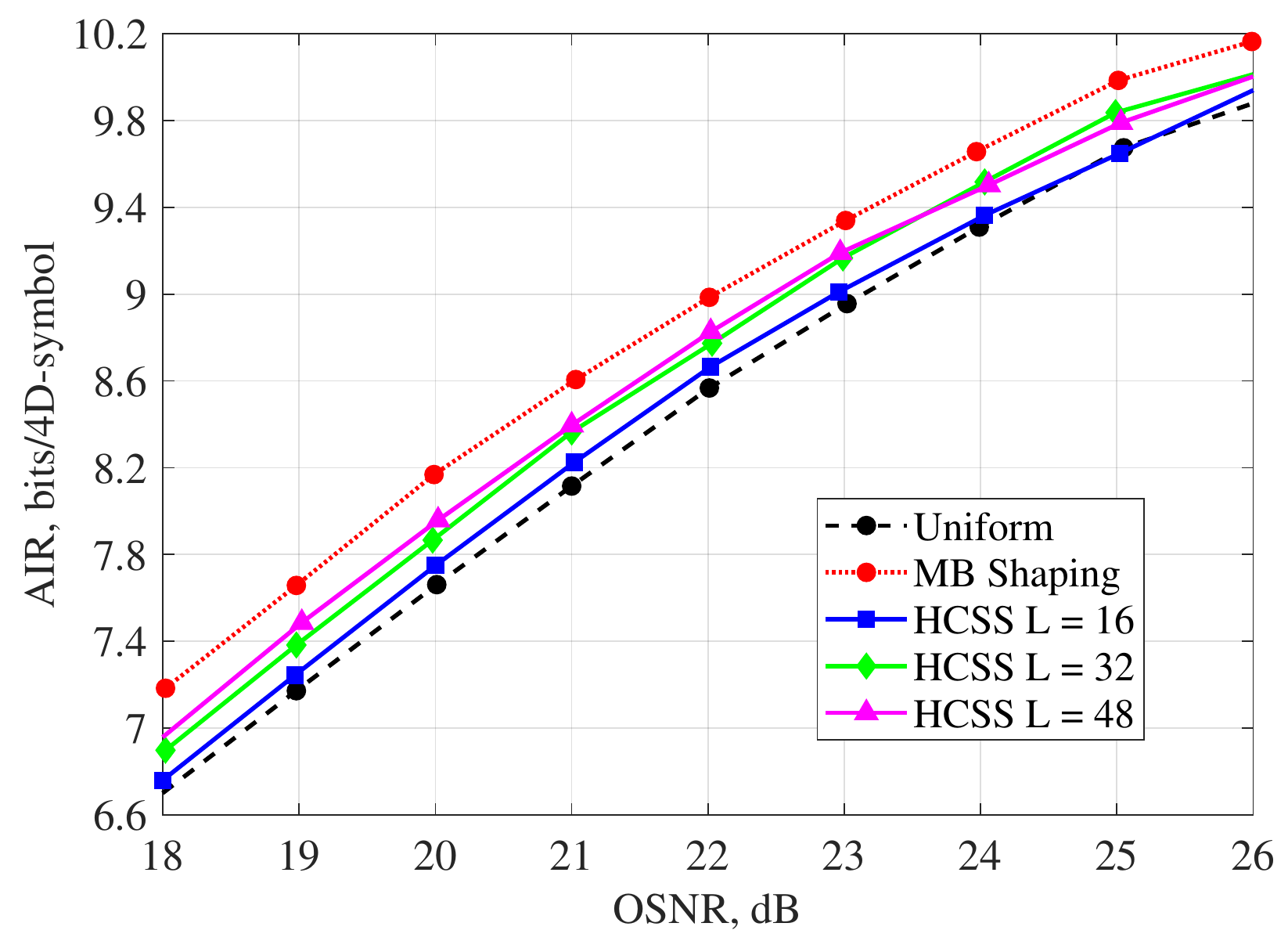}}\\
\subfloat[]{\includegraphics[scale = 0.50]{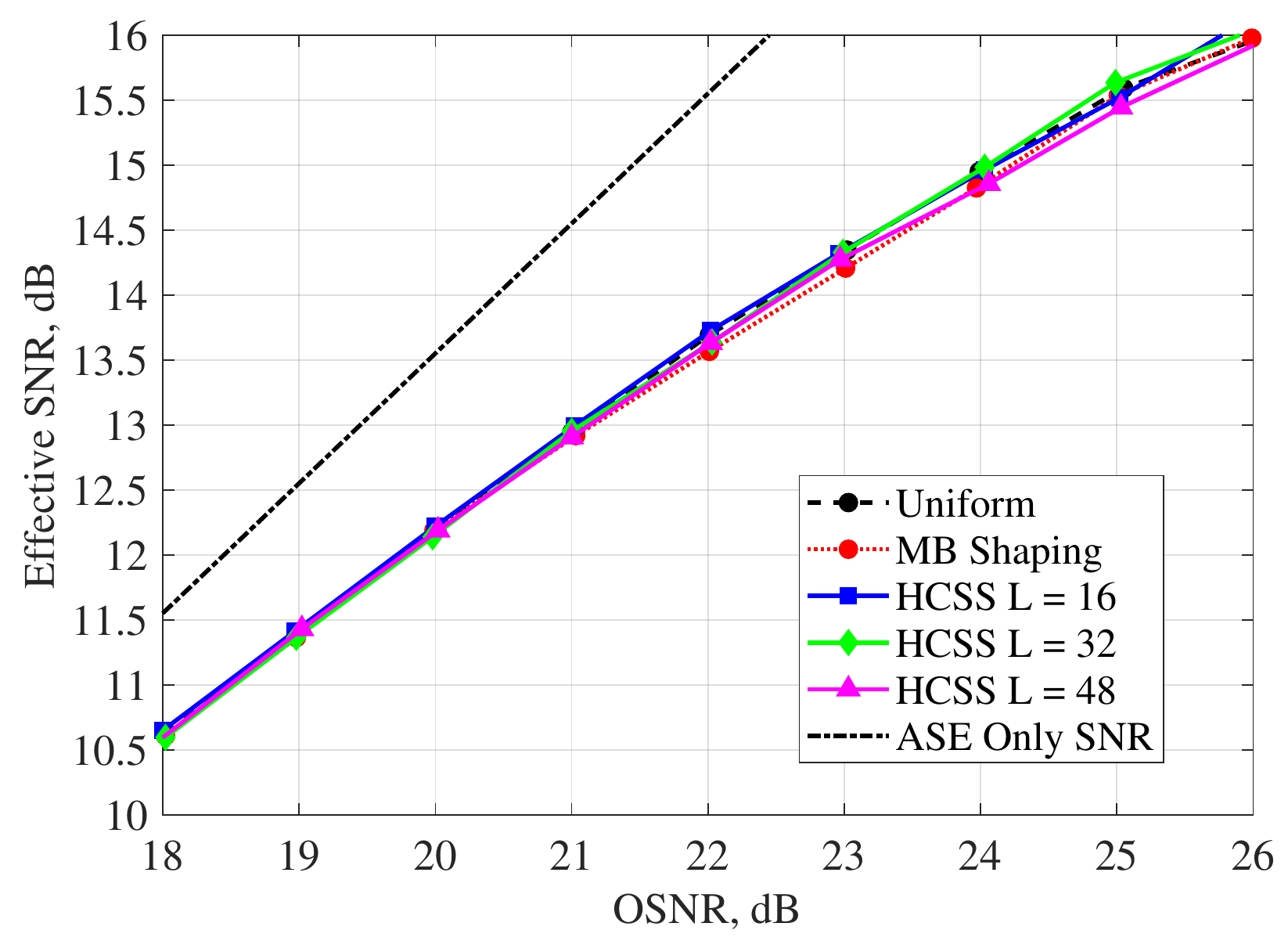}}
\caption{Performance vs. OSNR in back-to-back configuration ($R_{\mathrm{S}}=1.75$; $L=16,32,48$; 4D symbol mapping): (a) AIR, (b) Effective SNR.}
\label{fig:B2B}
\end{figure}

\begin{figure}[!t]
\centering
\captionsetup[subfloat]{farskip=1pt}
\subfloat[]{\includegraphics[width=0.15\textwidth]{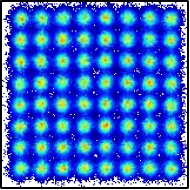}}
\hfil
\subfloat[]{\includegraphics[width=0.15\textwidth]{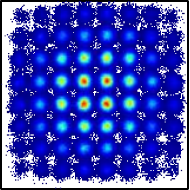}}
\hfil
\subfloat[]{\includegraphics[width=0.15\textwidth]{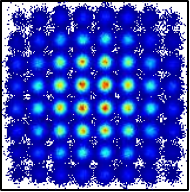}}
\caption{Experimental constellations BtB under high OSNR: (c) Uniform, (b) Maxwell--Boltzmann shaping, (c) Huffman-coded sphere shaping ($L=32$).}
\label{fig:Constellations}
\end{figure}

\begin{figure}[!t]
\centering
\captionsetup[subfloat]{farskip=1pt}
\subfloat[]{\includegraphics[scale=0.50]{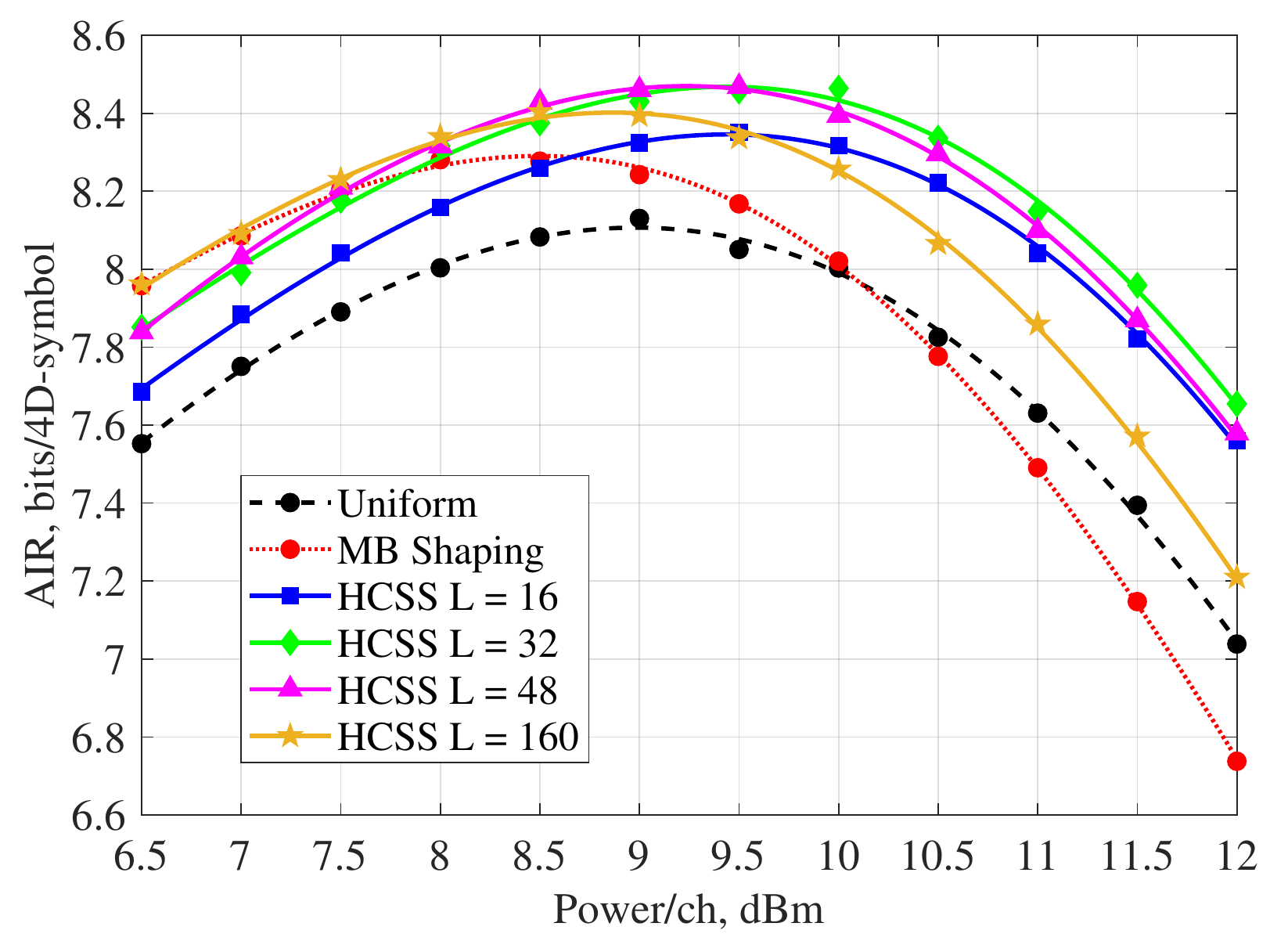}}\\
\subfloat[]{\includegraphics[scale=0.50]{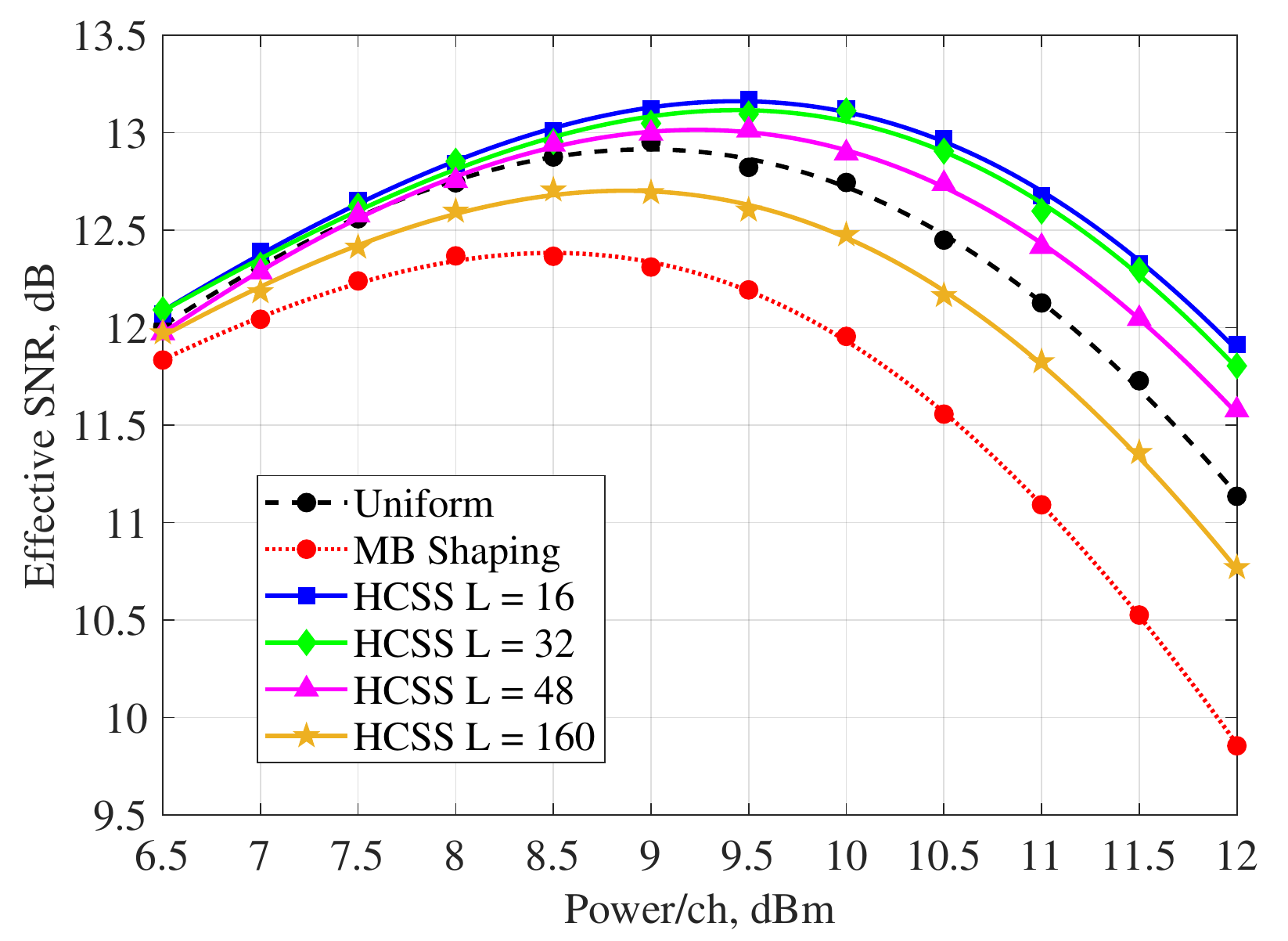}}
\caption{Performance vs. optical launch power for single-span transmission ($R_{\mathrm{S}}=1.75$; $L=16,32,48,160$; 4D symbol mapping): (a) AIR, (b) Effective SNR.}
\label{fig:PowerSweep_Length}
\end{figure}

\begin{figure}[!t]
\centering
\captionsetup[subfloat]{farskip=1pt}
\subfloat[]{\includegraphics[scale=0.50]{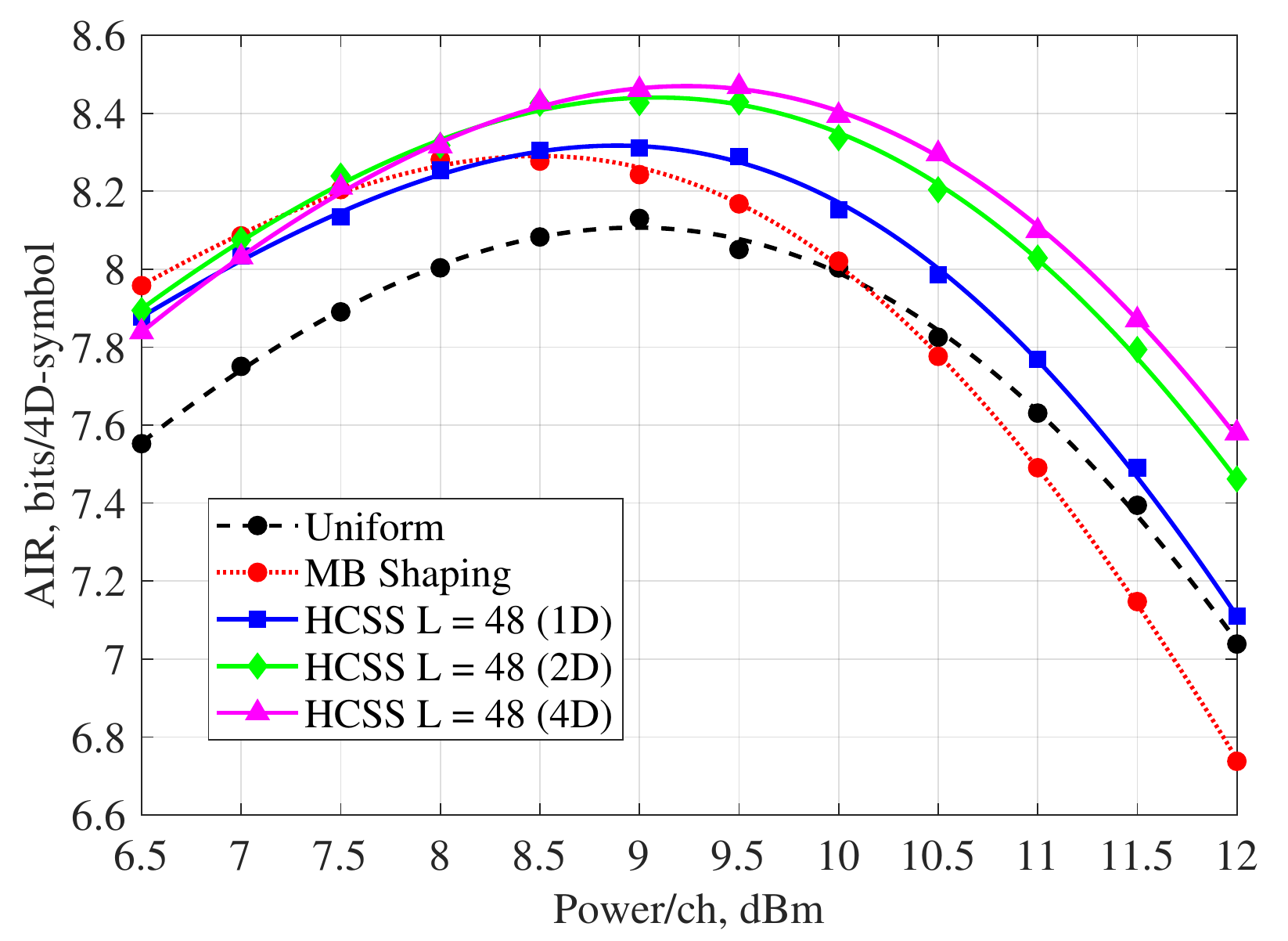}}\\
\subfloat[]{\includegraphics[scale=0.50]{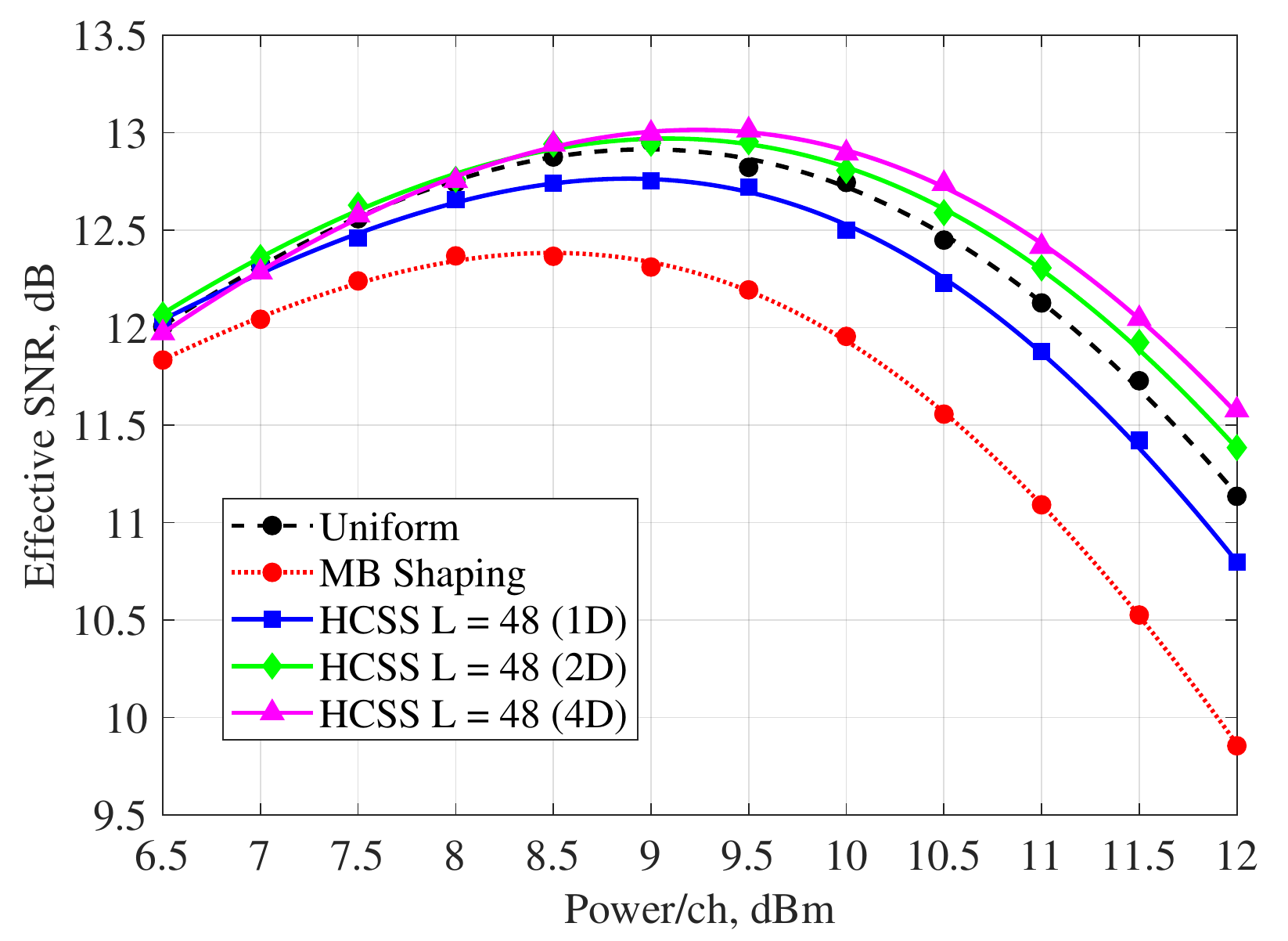}}
\caption{Performance vs. optical launch power for single-span transmission ($R_{\mathrm{S}}=1.75$; $L=48$; 1D, 2D and 4D symbol mapping): (a) AIR, (b) Effective SNR.}
\label{fig:PowerSweep_Dim}
\end{figure}


\begin{figure*}[!t]
\centering
\captionsetup[subfloat]{farskip=1pt}
\subfloat[]{\includegraphics[scale=0.50]{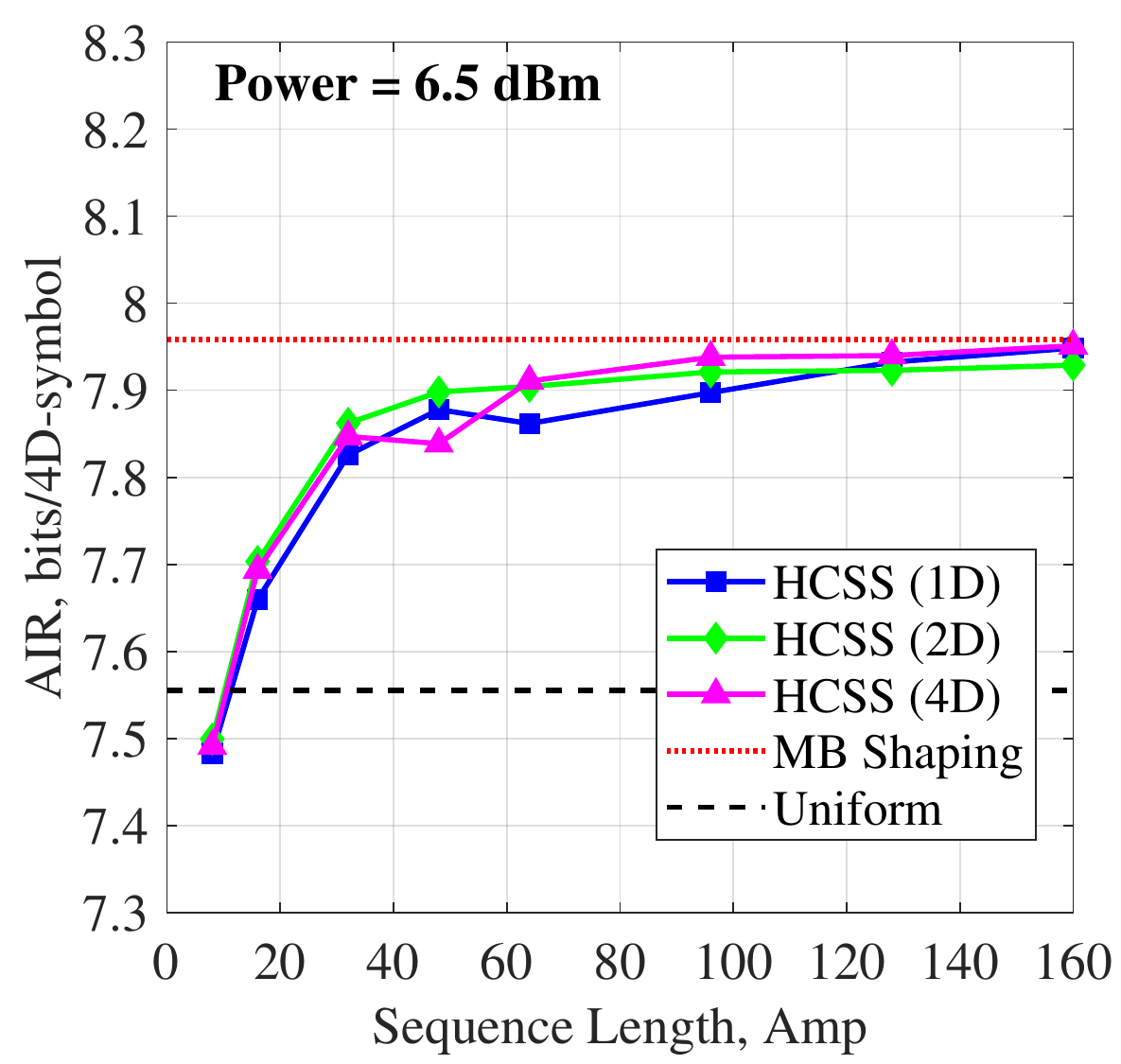}}
\subfloat[]{\includegraphics[scale=0.50]{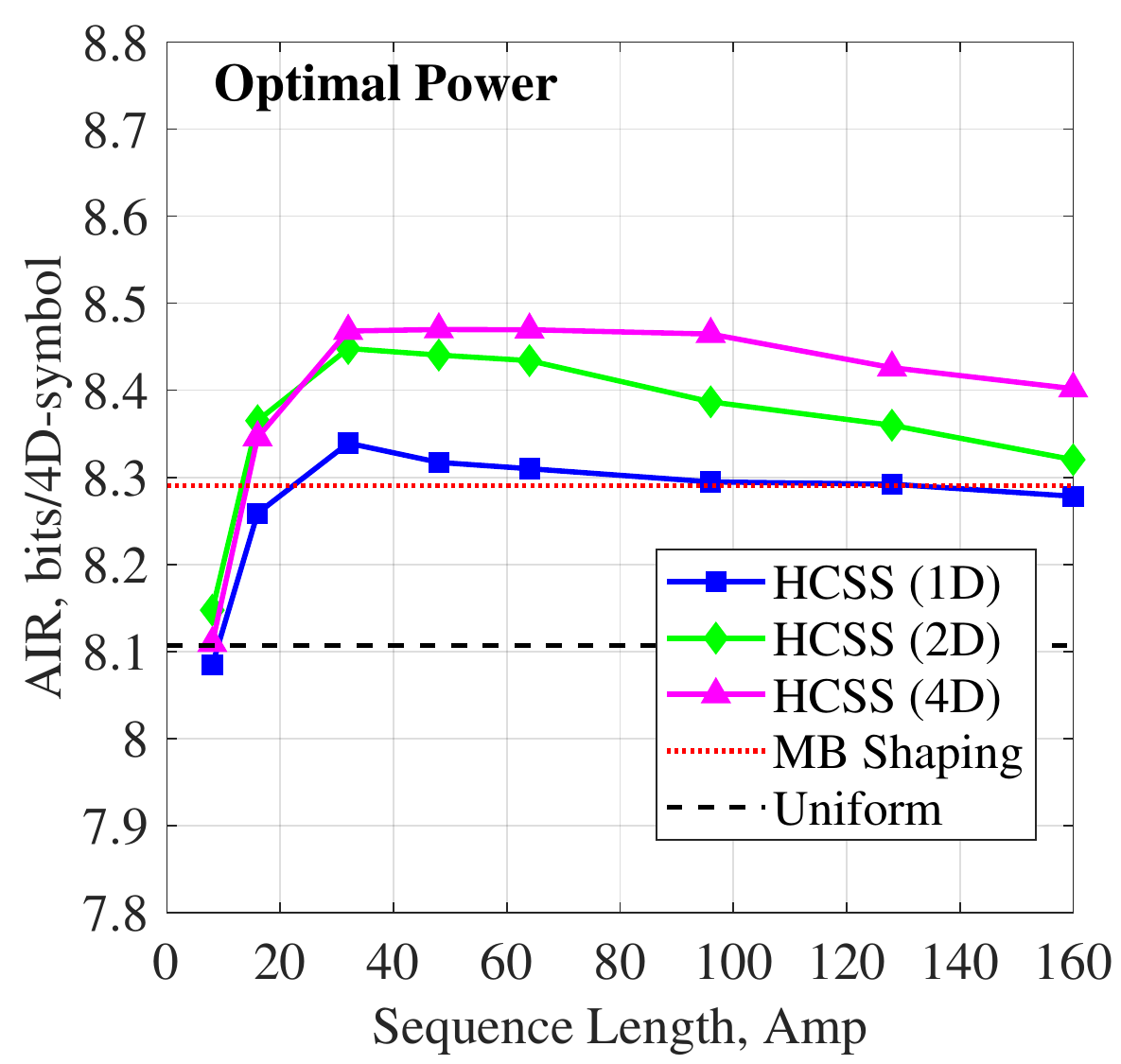}}
\subfloat[]{\includegraphics[scale=0.50]{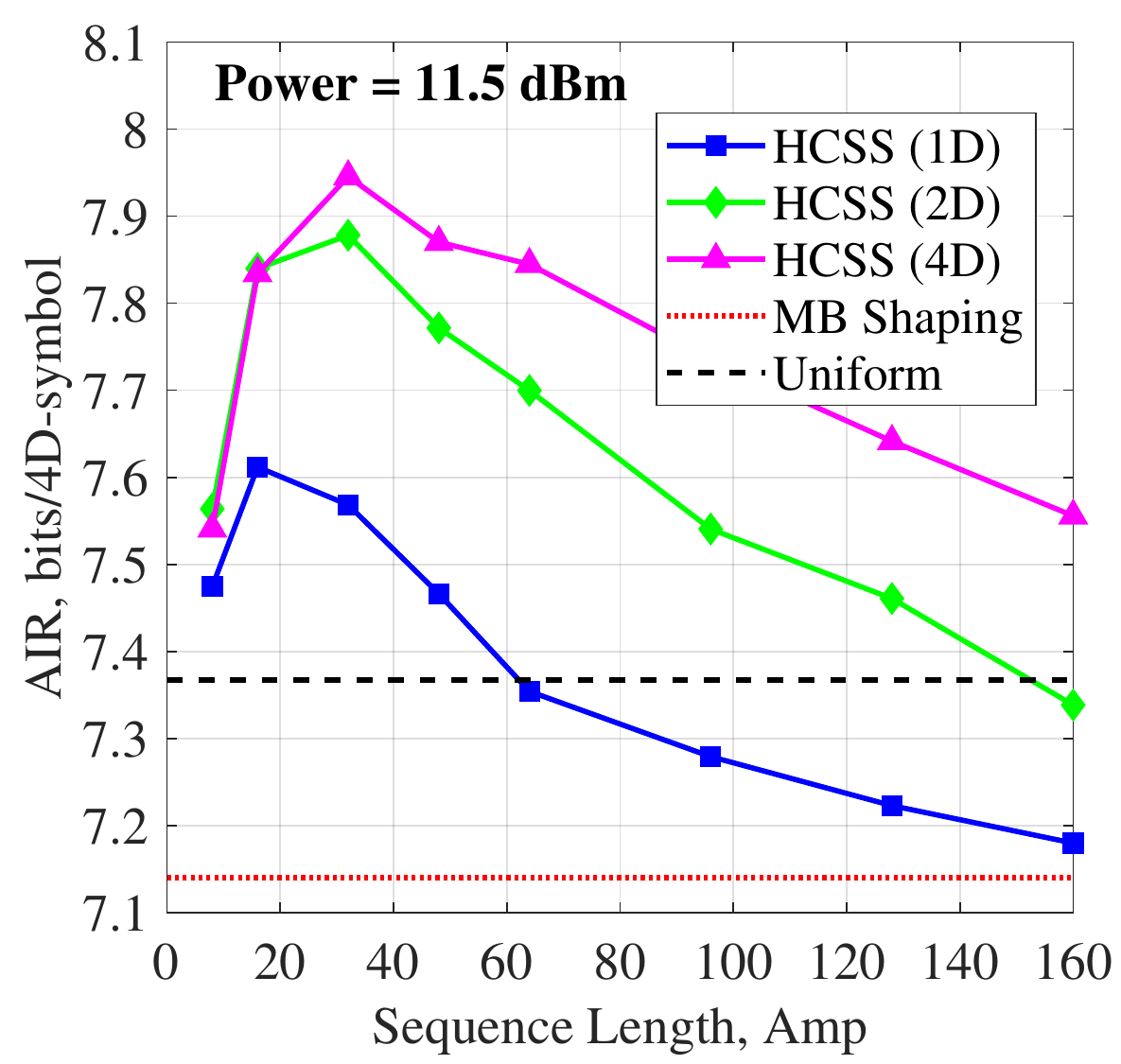}}\\

\subfloat[]{\includegraphics[scale=0.50]{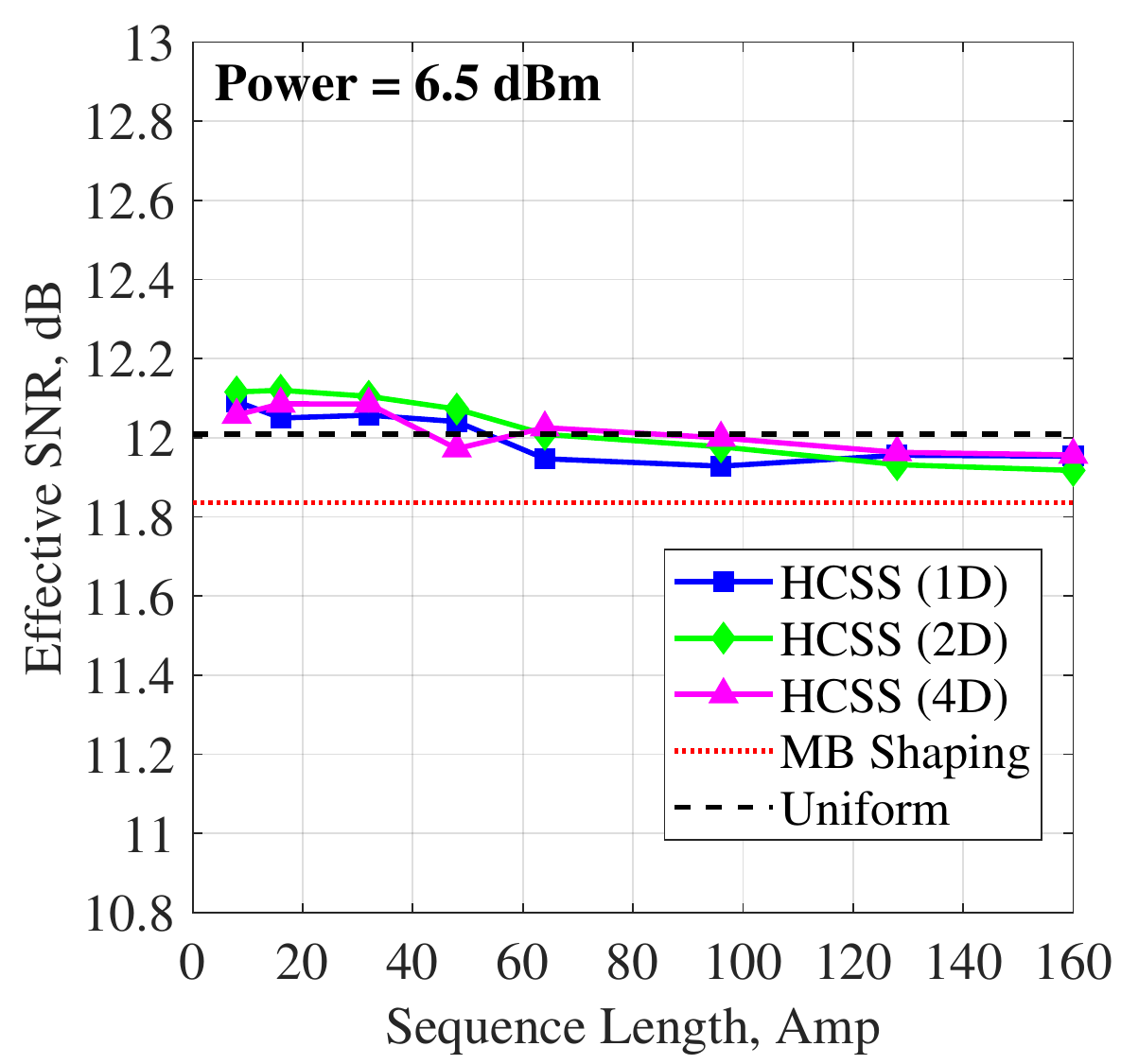}}
\subfloat[]{\includegraphics[scale=0.50]{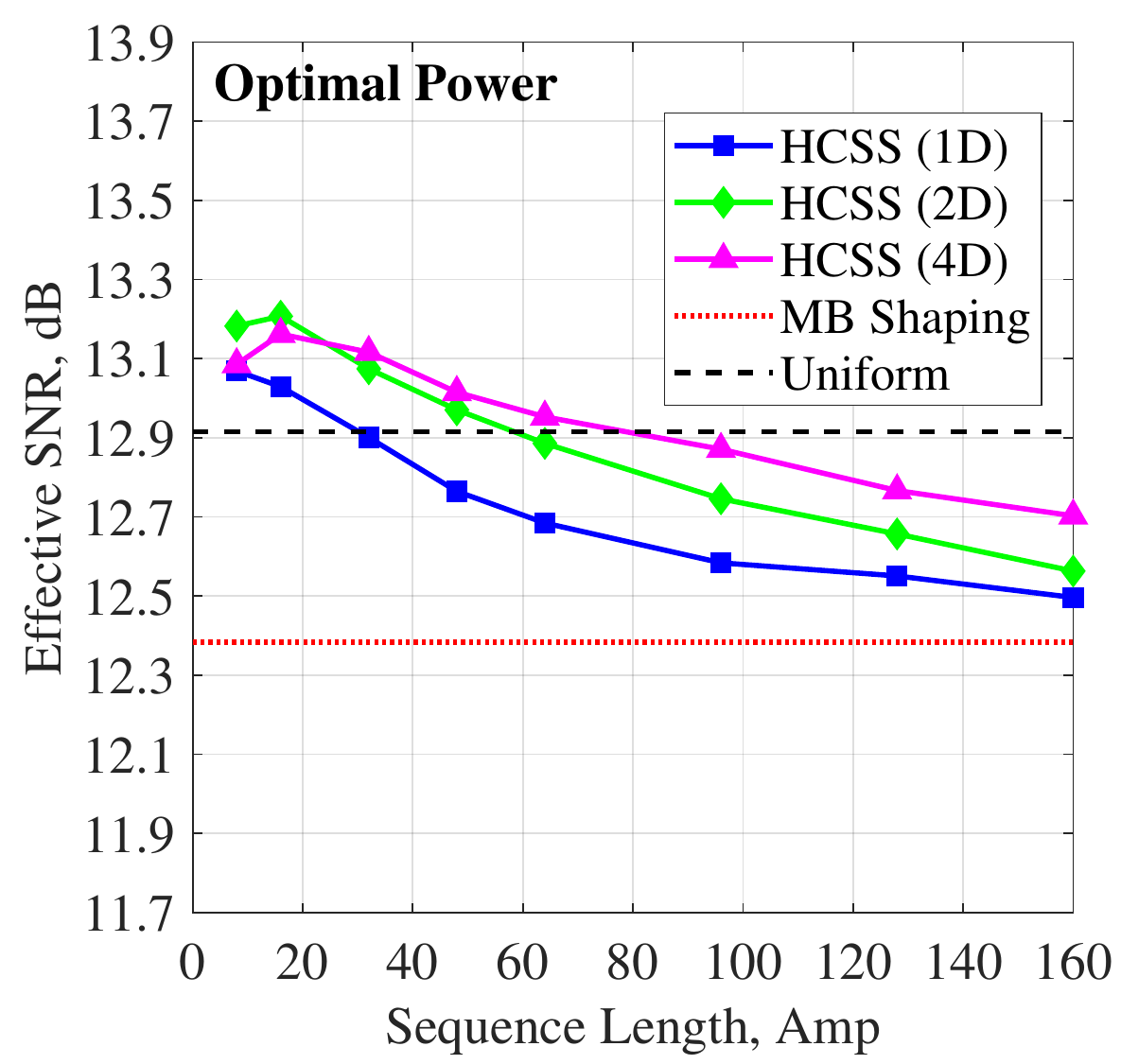}}
\subfloat[]{\includegraphics[scale=0.50]{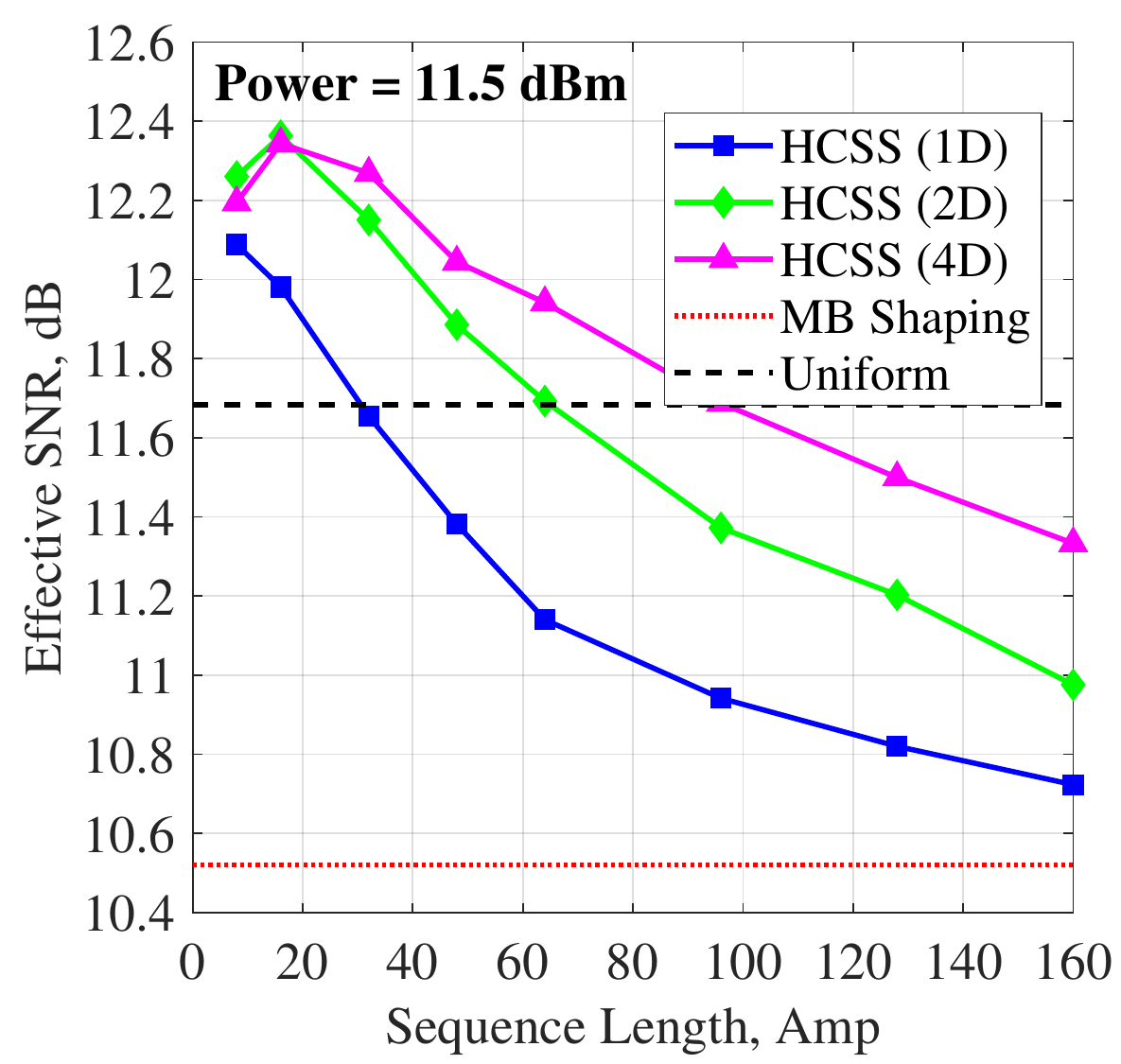}}
\caption{Performance vs. shaping sequence length ($R_{\mathrm{S}}=1.75$; 1D, 2D and 4D symbol mapping): (a) and (d) AIR and effective SNR for launch power of 6.5~dBm, (b) and (e) AIR and effective SNR for optimal launch power, (c) and (f) AIR and effective SNR for launch power of 11.5~dBm.}
\label{fig:OptimalLength}
\end{figure*}

\begin{figure}[!t]
\centering
\captionsetup[subfloat]{farskip=1pt}
\subfloat[]{\includegraphics[scale=0.50]{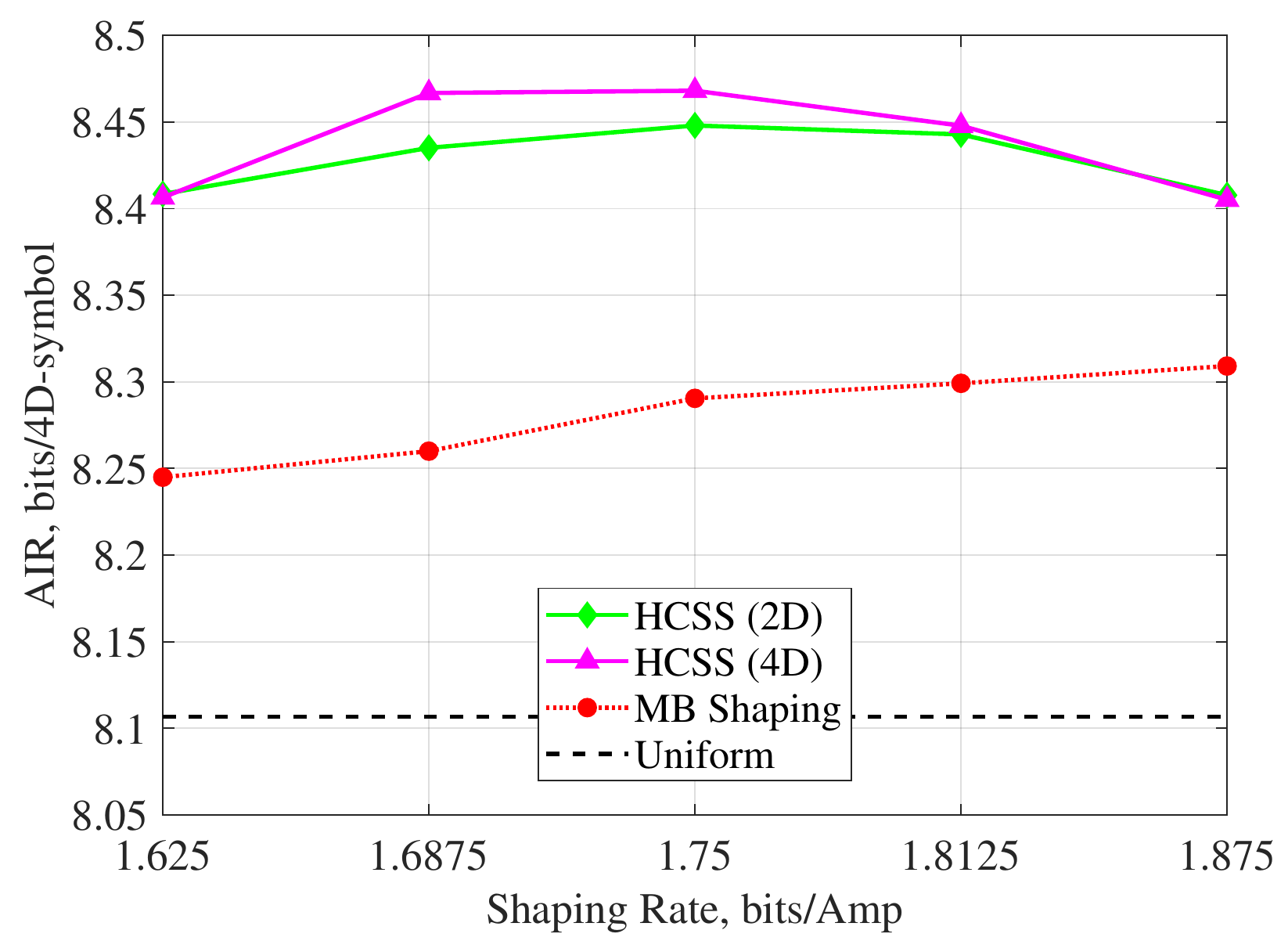}}\\
\subfloat[]{\includegraphics[scale=0.50]{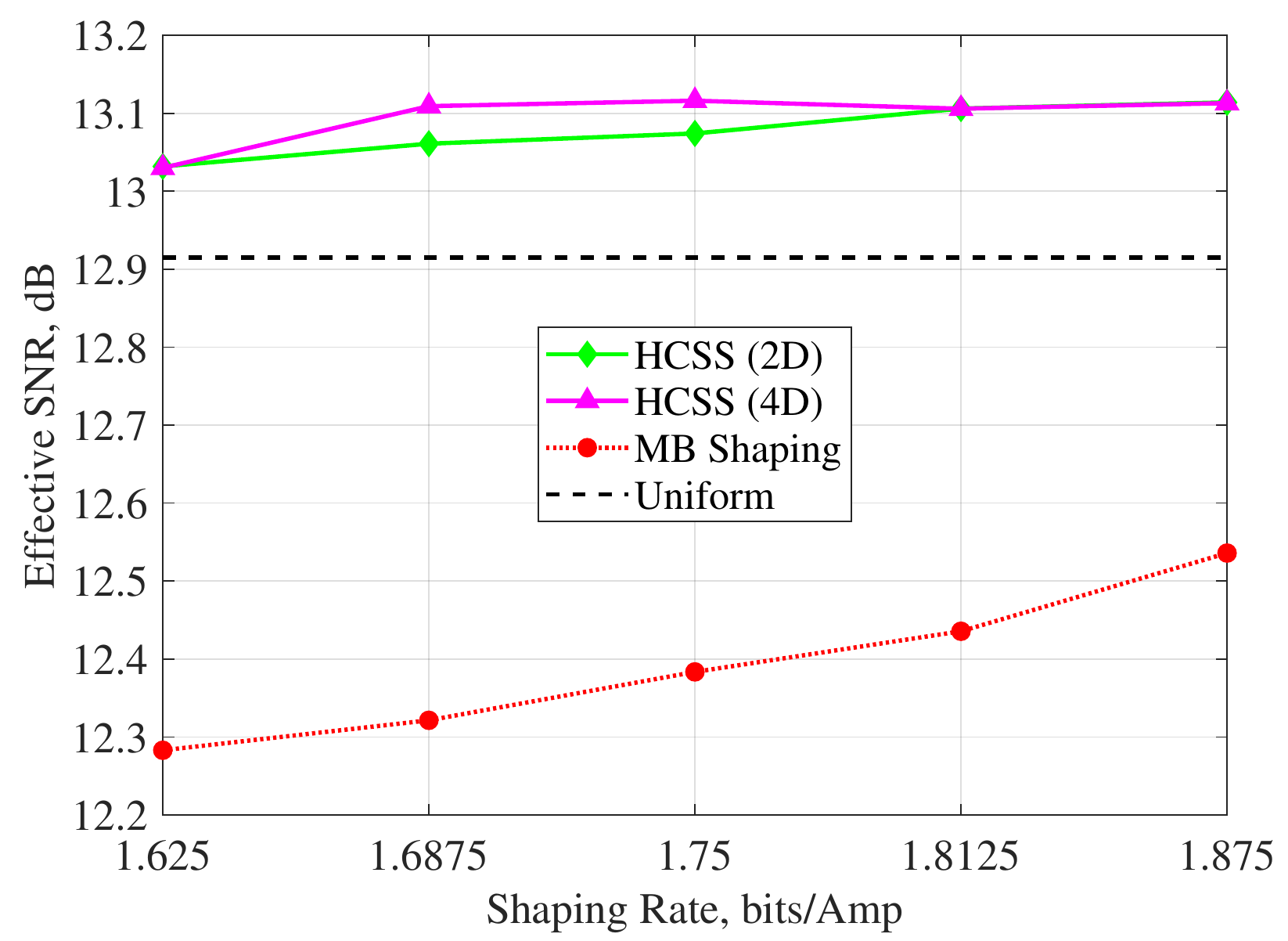}}
\caption{Performance vs. shaping rate ($L = 32$; 2D and 4D symbol mapping): (a) AIR, (b) Effective SNR.}
\label{fig:OptimalRate}
\end{figure}

\subsection{Back-to-back}

Fig.~\ref{fig:B2B} shows a back-to-back characterization of the system under consideration. We compared uniform signaling, MB shaping at $R_{\mathrm{S}} = 1.75$~b/Amp, and HCSS with $L=16,32,48$  using 4D symbol mapping at the same shaping rate. Fig.~\ref{fig:B2B}(a) illustrates the variation of AIR with OSNR. We observe that infinite-length MB shaping achieves superior performance with a $0.48$~b/4D AIR gain compared with uniform signaling over an operating OSNR range of $18$--$21$~dB. HCSS achieves gains of $0.08$, $0.22$ and $0.32$~b/4D for $L = 16, 32, 48$, respectively. This shows the trend that longer length shaping achieves better performance (eventually, approaching that of MB shaping), as is expected for AWGN channels. We note that for higher OSNR values, shaping gain reduces both for HCSS and MB shaping.    

Fig.~\ref{fig:B2B}(b) illustrates the variation of effective SNR with OSNR. We observe that there is no noticeable difference in terms of effective SNR over an operating OSNR range of $18$--$21$~dB (SNR range of $10.5$--$13$~dB) for the shaping schemes under consideration. Therefore, we conclude that shaping algorithms do not introduce additional implementation penalty over uniform signaling over the range of SNRs under consideration. We also note that shaping sequence length and symbol mapping strategy do not impact the SNR in back-to-back.  

Fig.~\ref{fig:Constellations} (a)--(c) shows constellation diagrams for uniform signaling, MB shaping and HCSS ($L=32$, 4D amplitude-to-symbol mapping) at $R_{\mathrm{S}} = 1.75$~b/Amp under high OSNR (about $35$~dB). We observe no notable visual difference between HCSS and MB shaped constellations.  

\subsection{Single-span transmission}

For analysis of the experimental data for single-span transmission, data fitting based on a Gaussian noise (GN)-model \cite{GN_model} was performed. A detailed explanation of the data fitting approach can be found in Appendix~\ref{SecondAppendix}.

\subsubsection{Optical launch power sweep}

Figs.~\ref{fig:PowerSweep_Length} and \ref{fig:PowerSweep_Dim} demonstrate system performance characterization in terms of AIR and effective SNR as a function of optical launch power for uniform signaling, MB shaping and HCSS at $R_{\mathrm{S}} = 1.75$~b/Amp. Individual data points represent measured experimental data, while corresponding smooth lines represent the data fit. We note that the model described in Appendix~\ref{SecondAppendix} provides a very good fit to the measured experimental data.

Fig.~\ref{fig:PowerSweep_Length} (a) and (b) show AIR and effective SNR for uniform signaling, MB shaping and HCSS using $L=16,32,48,160$ with 4D symbol mapping. In the linear regime, performance is consistent with that of the back-to-back measurements. HCSS with longer sequence length achieves higher AIR (approaching the performance of MB shaping with $L=160$), while variation in effective SNR is relatively small. At the optimal launch power, we observe that MB shaping achieves a gain over uniform signaling of $0.18$~b/4D, while HCSS exhibits gains of $0.24, 0.37, 0.38, 0.30$~b/4D for $L=16, 32, 48, 160$, respectively. MB shaping suffers from severe nonlinear impairment, which can be seen as effective SNR degradation of $0.52$~dB compared to uniform signaling at optimum power, while HCSS demonstrated improved nonlinearity tolerance. HCSS with $L=16,32,48$ elicits SNR gains of $0.3$, $0.17$, $0.06$~dB compared to uniform signaling at optimal power, while for $L=160$ the effective SNR is reduced by $0.2$~dB. In the highly nonlinear regime (launch power above $\sim~11$~dBm), the performance degradation for MB shaping and HCSS with $L=160$ is more significant, whereas the gain for HCSS with shorter shaping length is increased.  

Fig.~\ref{fig:PowerSweep_Dim} (a) and (b) show AIR and effective SNR for uniform signaling, MB shaping and HCSS using $L=48$ with 1D, 2D and 4D symbol mapping strategies. We observe that HCSS with higher-dimensional symbol mapping achieves better performance in the nonlinear regime. At the optimal launch power, HCSS with 4D symbol mapping achieves AIR gains of $0.04$, $0.15$~b/4D with corresponding SNR gain of $0.05$, and $0.25$~dB compared to 2D and 1D symbol mapping, respectively. In the linear regime, the performance difference is negligible among all symbol mapping strategies.

\subsubsection{Optimal shaping length}

In Fig.~\ref{fig:OptimalLength} we examine the performance in terms of AIR and effective SNR when varying the shaping sequence length of HCSS ($L=64,96,128$ are added into consideration) with 1D, 2D and 4D symbol mapping strategies in linear, optimal launch power and highly nonlinear regimes. We note that the results shown in Fig.~\ref{fig:OptimalLength} are based on fitting of launch power sweep measurements.

As discussed previously, in the linear regime, which is shown in Fig.~\ref{fig:OptimalLength} (a) and (d), HCSS with longer shaping sequence length provides higher AIR and closely approaches MB shaping performance with $L=160$. No significant difference is observed among all symbol mapping strategies. Effective SNR also varies insignificantly --- there is minor SNR gain at shorter sequence lengths due to weak presence of nonlinearities.   

In the optimal launch power regime, which is shown in Fig.~\ref{fig:OptimalLength} (b) and (e), HCSS using 2D and 4D symbol mapping demonstrates significant performance gain in terms of AIR with the shaping sequence length $L$ in the range of $32$--$96$, while HCSS using 1D symbol mapping provides performance close to MB shaping (with $L\geq16$). The sequence length $L=32$ can be considered optimal (achieving the highest AIR) for all symbol strategies --- AIR gain over uniform signaling is $0.37$, $0.34$, $0.22$~b/4D (and $0.19$, $0.16$, $0.06$~b/4D over MB shaping) for 4D, 2D and 1D symbol mapping, respectively. AIR gain is supported by the improvement in effective SNR --- 4D mapping provides the highest SNR gain compared to 1D and 2D mapping with $L\geq32$, while 2D mapping slightly outperforms with $L\leq16$. For optimal sequence length of $L=32$, SNR gain over uniform signaling is $0.2$, $0.16$, $0$~dB (and $0.72$, $0.68$, $0.52$~dB over MB shaping) for 4D, 2D and 1D symbol mapping, respectively. 

In the highly nonlinear regime, which is shown in Fig.~\ref{fig:OptimalLength} (c) and (f), a similar performance trend can be observed as for optimal launch power regime, however, the gain (both in AIR and SNR) with short shaping length is exaggerated. For instance, $0.58$~b/4D AIR gain and $0.6$~dB SNR gain over uniform signaling ($0.83$~b/4D and $1.75$~dB over MB shaping) can be observed with $L=32$ and 4D symbol mapping.

In general, we note that HCSS with longer shaping sequence length can achieve better power efficiency for a fixed rate, and therefore provide higher shaping gain in a linear channel. This is clearly seen in the back-to-back case and linear regime of single-span transmission. In the presence of significant fiber nonlinearities, shorter shaping sequences provide higher nonlinear tolerance by improving the effective SNR in the received signal. By choosing appropriate shaping length for the operating regime, an optimal trade-off of shaping gain and effective SNR gain can be achieved. Also, increasing the dimensionality of the symbol mapping can improve nonlinear performance by shortening the shaped symbol sequence length in the time-domain and reducing probabilities of high peak power values, while maintaining the same power efficiency.   

\begin{figure}[!t]
\centering
\includegraphics[scale=0.50]{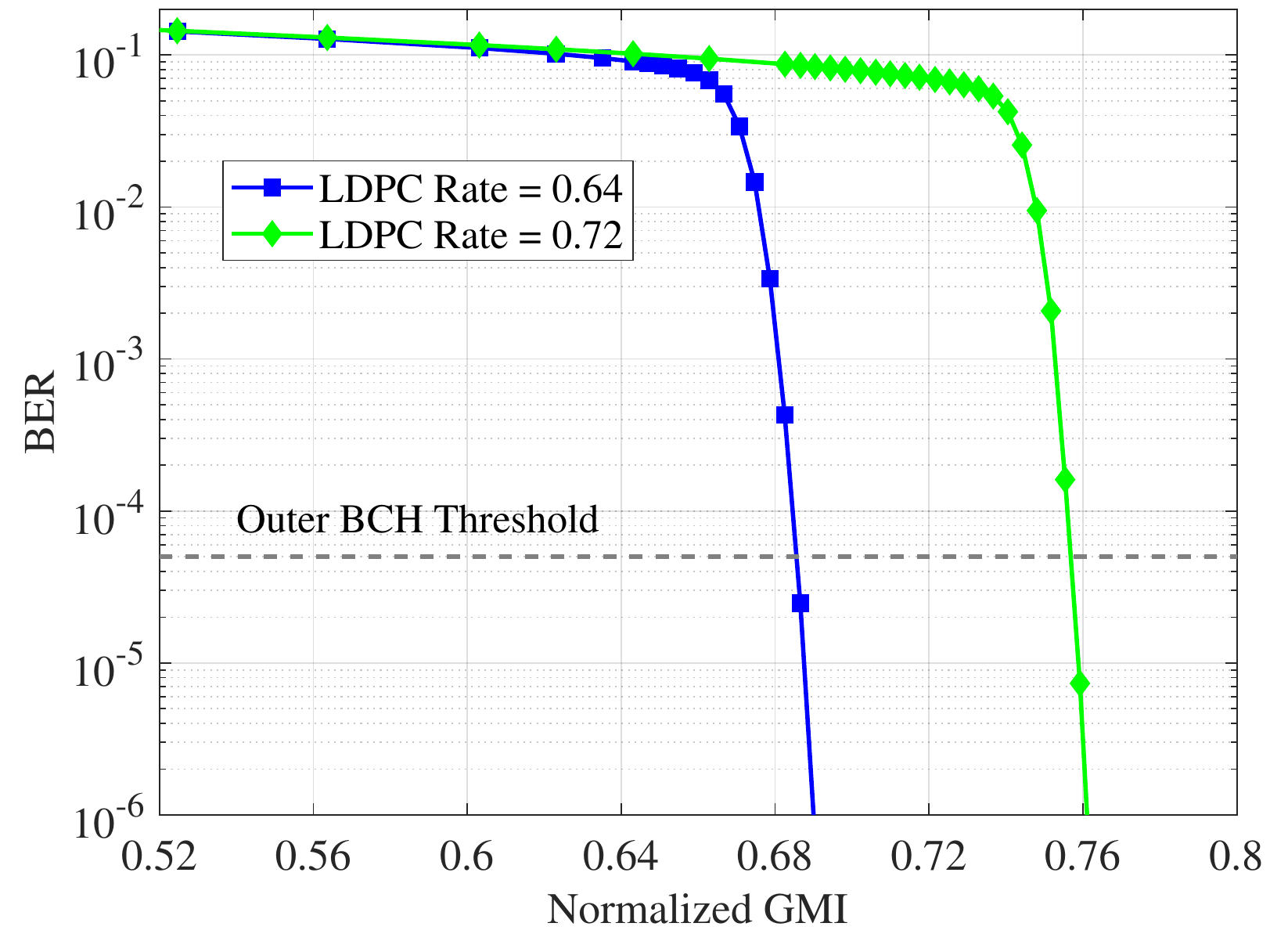}
\caption{Characterization of LDPC codes: BER vs. normalized GMI.}
\label{fig:LDPC}
\end{figure}

\subsubsection{Optimal shaping rate}
Next, we studied the impact of the shaping rate on the transmission performance. Fig.~\ref{fig:OptimalRate} shows a characterization of the AIR and effective SNR as a function of the shaping rate. For HCSS we considered shaping sequence length $L=32$ with 2D and 4D symbol mapping. Each data point is based on the data fit of launch power sweep measurements and represents the performance at the optimal launch power.

From Fig.~\ref{fig:OptimalRate} (a) we observe that $R_{\mathrm{S}} = 1.75$~b/Amp is the optimal shaping rate for HCSS --- AIR varies over $0.07$~b/4D in $R_{\mathrm{S}}$ range of $1.625$--$1.875$~b/Amp achieving the highest value at $R_{\mathrm{S}} = 1.75$~b/Amp, the corresponding variation in effective SNR is $0.1$~dB. For MB shaping, the AIR increases with the shaping rate by $0.06$~b/4D (within $R_{\mathrm{S}}$ of consideration), which is supported by the corresponding increase in effective SNR of $0.25$~dB.

In the case of HCSS with fixed shaping sequence length, the optimal effective SNR does not depend significantly on the shaping rate. Therefore, we may infer that the optimal shaping rate is mostly affected by the linear shaping gain contribution. Conversely, MB shaping demonstrates stronger dependence of both effective SNR and AIR on the shaping rate, indicating that the increase in AIR with shaping rate is associated with nonlinear transmission gain. We note that with increasing shaping rate, MB shaping will converge to uniform signaling, while HCSS will exhibit some rate loss due to the dyadic distribution of compositions constraint.


\begin{figure}[!t]
\centering
\captionsetup[subfloat]{farskip=1pt}
\subfloat[]{\includegraphics[scale=0.50]{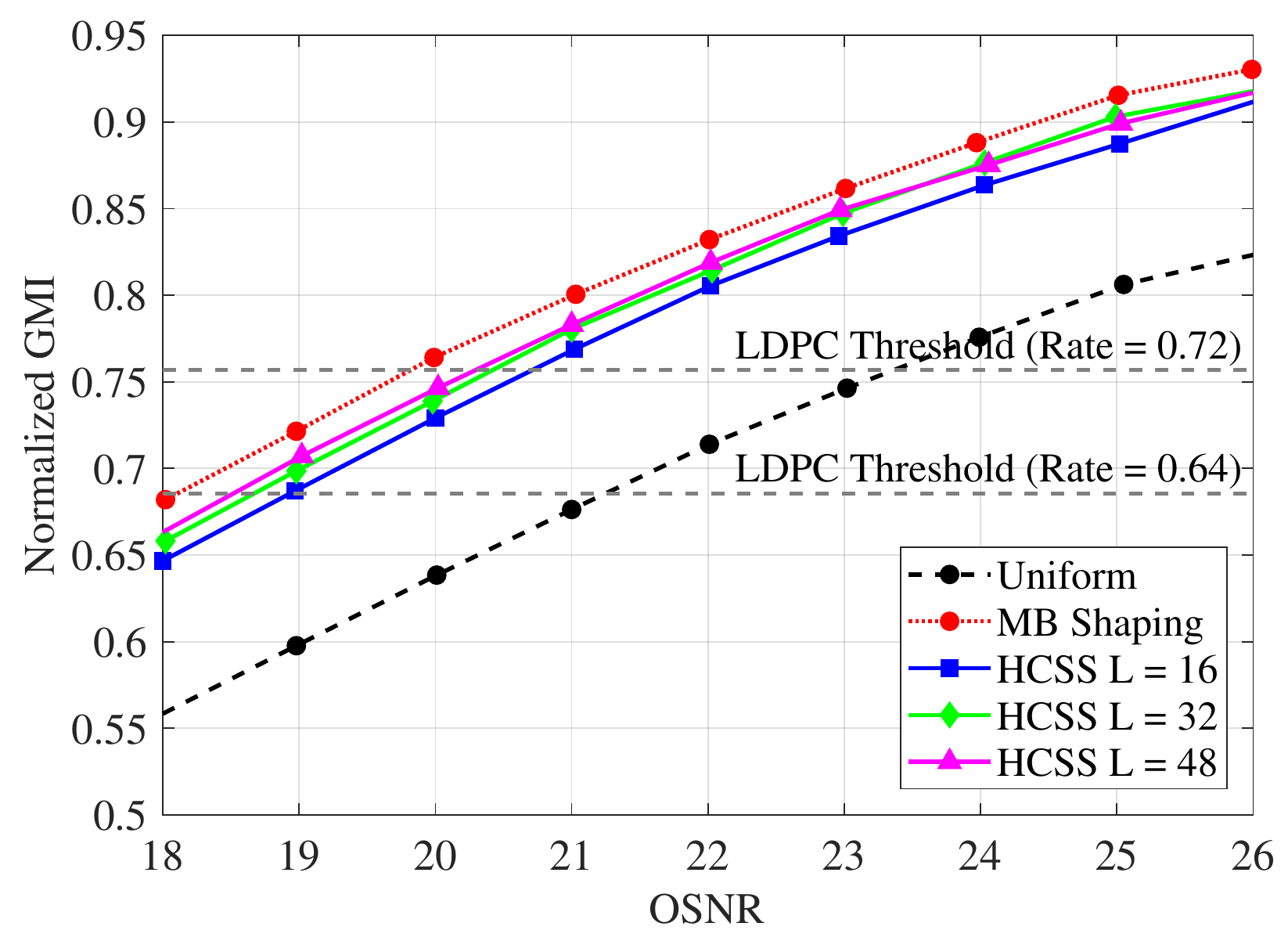}}\\
\subfloat[]{\includegraphics[scale=0.50]{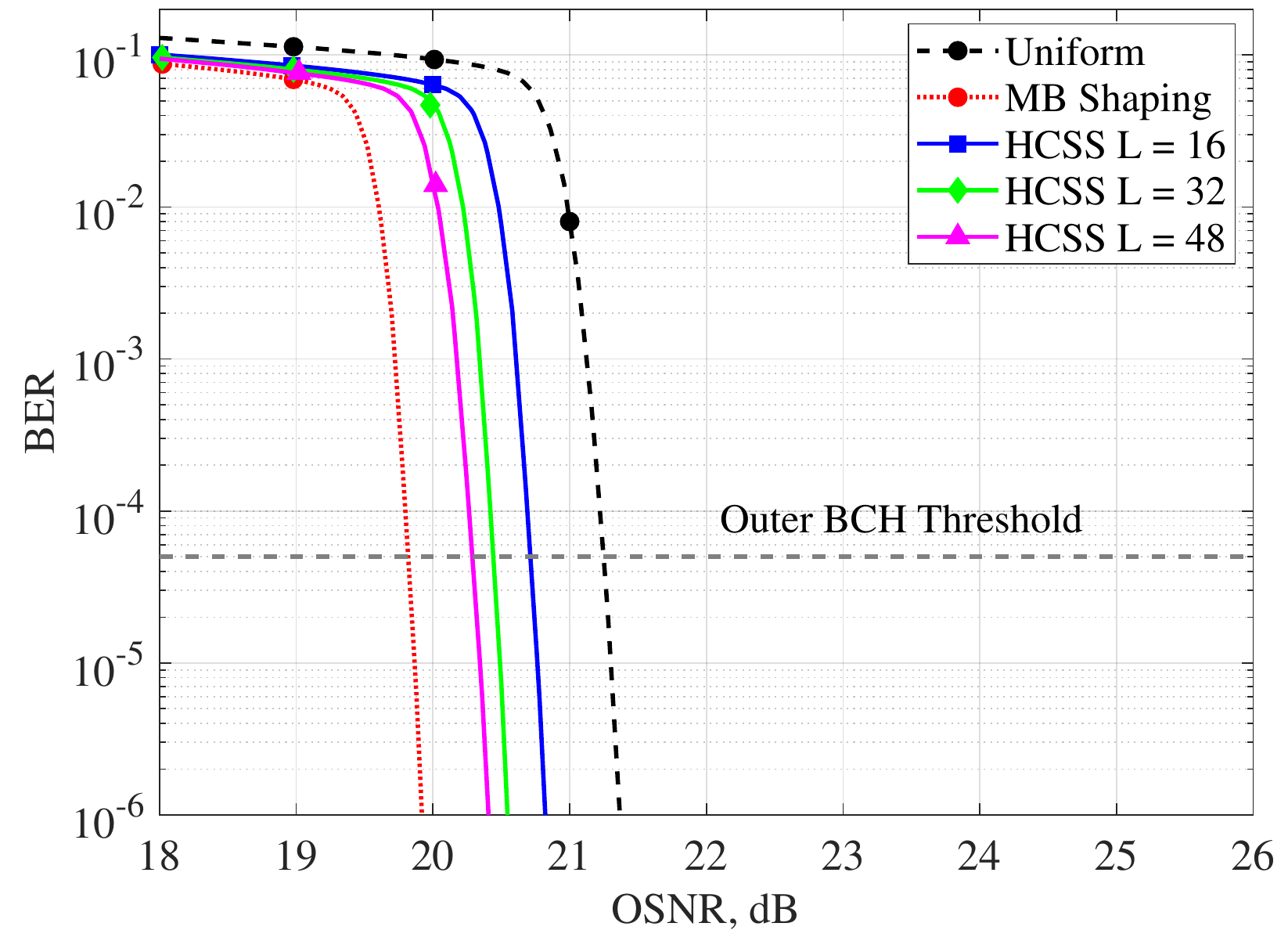}}
\caption{Coded performance analysis in back-to-back configuration: (a) normalized GMI vs. OSNR, (b) predicted BER after LDPC vs. OSNR.}
\label{fig:coded_btb}
\end{figure}

\begin{figure}[!t]
\centering
\captionsetup[subfloat]{farskip=1pt}
\subfloat[]{\includegraphics[scale=0.50]{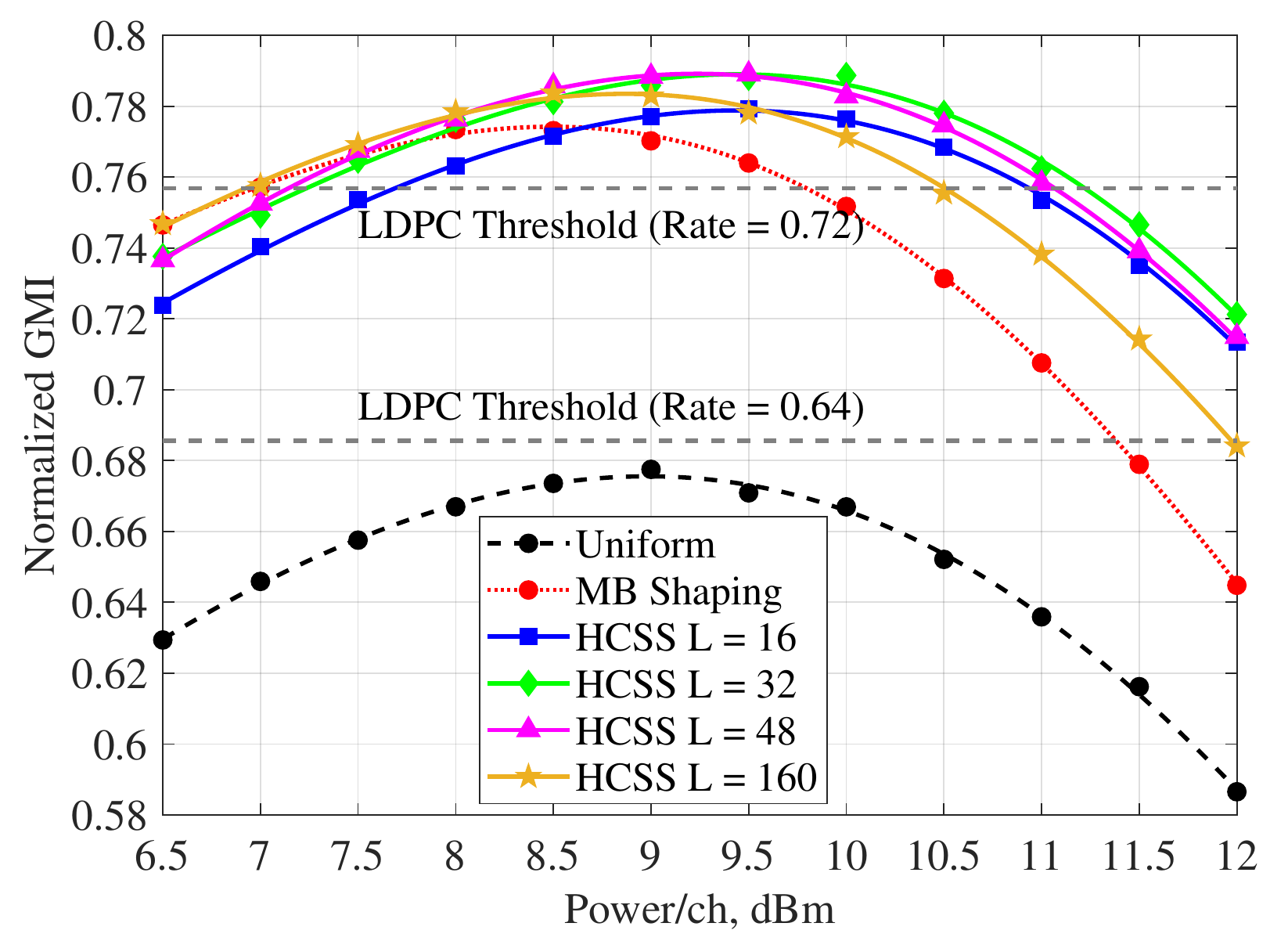}}\\
\subfloat[]{\includegraphics[scale=0.50]{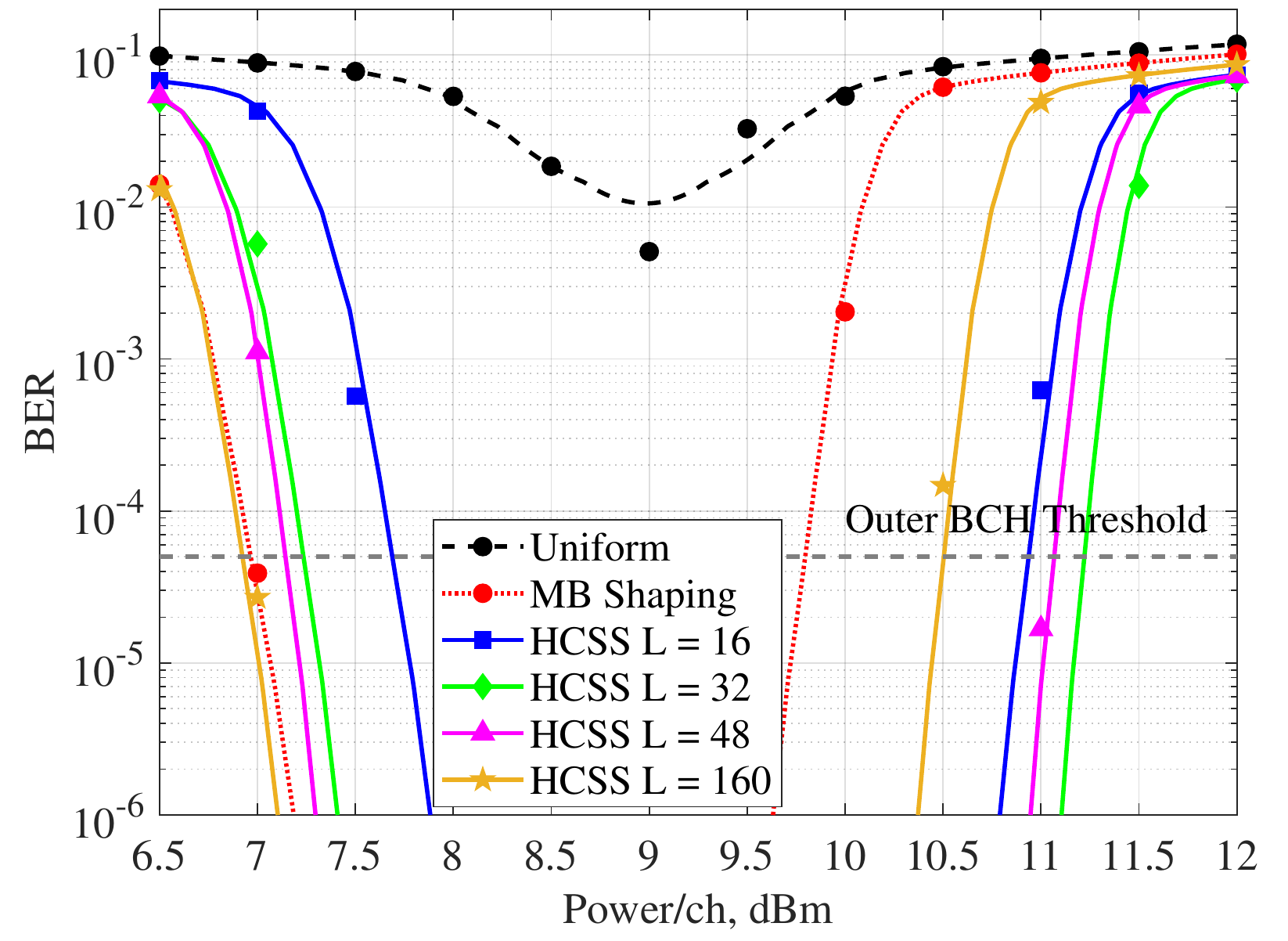}}
\caption{Coded performance analysis for single span transmission: (a) normalized GMI vs. optical launch power, (b) predicted BER after LDPC vs. optical launch power.}
\label{fig:coded_singlespan}
\end{figure}

\subsection{Coded performance}
For coded performance analysis we considered MB shaping and HCSS with 4D symbol mapping at $R_{\mathrm{S}} = 1.75$~b/Amp. The net bit rate after shaping and coding was $424.1$~Gb/s for both MB shaping and HCSS, while for uniform signaling the net rate was $426.7$~Gb/s. Bit error ratio (BER) after LDPC decoding (Fig.~\ref{fig:coded_btb} (b) and Fig.~\ref{fig:coded_singlespan} (b)) is predicted based on $\textrm{nGMI}$ (Fig.~\ref{fig:coded_btb} (a) and Fig.~\ref{fig:coded_singlespan} (a)). The individual points represent predicted BER based on measured nGMI values, while corresponding smooth lines represent predicted BER based on fitted nGMI values (which are calculated from AIR fit using (\ref{ngmi_uni}) and (\ref{ngmi_shaped})). 

The characterization of LDPC codes in terms BER as a function of $\textrm{nGMI}$ is shown in Fig.~\ref{fig:LDPC}. The LDPC $\textrm{nGMI}$ thresholds for the outer BCH code (such that the BER at the output of LDPC code is below the BCH code threshold of $5\times 10^{-5}$) are $0.757$ and $0.686$ for code rates of $0.72$ and $0.64$, respectively.  

Fig.~\ref{fig:coded_btb}  demonstrates coded performance for the back-to-back configuration. MB shaping achieves the best sensitivity, while HCSS with longer shaping length achieves better sensitivity than with shorter length --- for MB shaping OSNR margin is improved by $1.3$~dB, for HCSS the OSNR margin is improved by $0.5$, $0.7$, $0.7$~dB for $L=16, 32, 48$, respectively. 

Coded performance for single-span transmission is shown in Fig.~\ref{fig:coded_singlespan}. In the linear regime the sensitivity shows the similar trend as for back-to-back configuration, while in nonlinear regime $L=32$ demonstrated the best sensitivity. With $L=32$ the launch power margin is improved by $1.3$~dB compared to the MB shaping. We note that in case of uniform signaling, desired performance is not achievable.

\section{Discussion and Conclusions}
\label{sec:conc}

We have investigated HCSS as a method for probabilistic constellation shaping for application in optical fibre communication systems comprising extended-reach single-span links, subject to strong nonlinearities. We have demonstrated that the na\"{i}ve approach of optimizing the signal PMF and attempting to achieve this with a distribution matcher, while being optimal in the AWGN channel, is highly suboptimal in this case, significantly reducing the ultimate shaping gain. Such a system can achieve only a maximum $0.18$~b/4D gain in $200$~km single-span SSMF links. Conversely, HCSS achieves a gain of $0.37$~b/4D with a shaping sequence length of only $L=32$ and 4D symbol mapping. Such an HCSS system can be implemented without multiplications, and LUT size of no more than $100$~kbit.

In extended-reach single-span transmissions, signals may suffer from strong highly correlated short-memory nonlinearities. Hence, the temporal structure of the signal (e.g., ordering of transmitted symbols) may have a significant impact on nonlinear performance. By shortening the shaping sequence length, it it possible to introduce some advantageous changes in the temporal structure --- e.g., reduced concentration of high-power symbols \cite{Fehenberger_ShortLength_CCDM}. Also, by increasing the dimensionality of symbol mapping, the effective shaped symbol sequence length can be reduced as well as the number of simultaneously interacting independent shaped sequences, which leads to reduced probability of high peak power values.

\appendices
\section{Probability Mass Function (PMF) Calculation}
\label{FirstAppendix}
To construct a 4D DP-64QAM format, we consider the alphabet of amplitudes $\mathcal{A}=\{a_1, a_2, a_3, a_4\}=\{1, 3, 5, 7\}$ and the alphabet of signs $\mathcal{S}=\{s_1,s_2\}=\{-1,1\}$. The alphabet of a 4D-signal (4D-constellation) can be expressed as the Cartesian product 
$\mathcal{X} = \mathcal{S}^4 \times \mathcal{A}^4$, where $\mathcal{S}^4$ and $\mathcal{A}^4$ are $4$-fold Cartesian products with themselves.

A composition is defined as $C_i = \{c_1^i,c_2^i,c_3^i,c_4^i\}$, where $i$ is the index of the composition in the Huffman-coded structure, and $c_k^i$ is the number of instances of amplitude $a_k$ in the shaped amplitude sequence of length $L=\sum_{k=1}^{4}c_k^i$.

We postulate that the probability distribution of signs $P_S(s)$ is uniform. Hence, $P_S(s) = 1/2$ for $\forall s \in \mathcal{S}$ and $P_{S^4}(s^4) = 1/16$ for $\forall s^4 \in  \mathcal{S}^4$. The PMF of a 4D-signal can be therefore expressed as   
\begin{equation}\label{pmf_total}
P_{X}(x)  = P_{S^4}(s^4) \cdot P_{A^4}(a^4) 
 = 1/16 \cdot P_{A^4}(a^4)\,,
\end{equation}
where $P_{A^4}(a^4)$ can be considered as the PMF of a 4D-quadrant and $a^4=[a_{k_1},a_{k_2},a_{k_3},a_{k_4}]$ is a 4D-amplitude vector. 

\subsection{1D symbol mapping}
Since all components of a 4D-amplitude vector are mapped from independent shaped amplitude sequences, the PMF for a 4D-quadrant is a product of 1D-PMFs:   
\begin{equation}\label{pmf1d_product}
P_{A^4}(a^4) = P_{A}(a_{k_1}) \cdot P_{A}(a_{k_2}) \cdot P_{A}(a_{k_3}) \cdot P_{A}(a_{k_4})\,.
\end{equation}
The 1D-PMF which results from the $i^\mathrm{th}$ composition can be calculated as $P_A^i(a_k) = c_k^i/L$, while the total 1D-PMF is 
\begin{equation}\label{pmf1d_total}
P_A(a_k) = \sum_{i=1}^N p_i P_A^i(a_k) = \sum_{i=1}^N p_i \frac{c_k^i}{L}\,,
\end{equation}
where $p_i$ is the probability of occurrence of the $i^\mathrm{th}$ composition, and $N$ is the total number of compositions. 

\subsection{2D symbol mapping}
In the case of 2D symbol mapping, two consecutive amplitudes from two independent shaped amplitude sequences are used to construct a 4D-amplitude vector. Therefore, the 4D-quadrant PMF can be expressed as the product of 2D-PMFs:  
\begin{equation}\label{pmf2d_product}
P_{A^4}(a^4) = P_{A^2}(a^2) \cdot P_{A^2}(a^{2\ast})\,,
\end{equation}
where $a^2=[a_{k_1},a_{k_2}]$ and $a^{2*}=[a_{k_3},a_{k_4}]$ are 2D-amplitude vectors, $a^4=[a^2,a^{2*}]$.

Since components of 2D-amplitude vectors are mapped from a single sequence and not independent, the 2D-PMF resulting from the $i^\mathrm{th}$ composition can be calculated as
\begin{equation}\label{pmf2d_composition}
P_{A^2}^i(a^2) =
\bigg[\frac{c_{k_1}^i}{L} \cdot \frac{c_{k_2}^i-\delta}{L-1} \bigg]^{+}\,,  
\end{equation}
where $[\cdot]^+$ denotes $\mathrm{max}\{\cdot,0\}$ operator, $\delta$ is 
\begin{equation}\label{pmf2d_correction}
\delta =
\begin{cases}
0, & k_1 \neq k_2, \\
1, & k_1 = k_2\,. 
\end{cases}
\end{equation}
The total 2D-PMF is
\begin{equation}\label{pmf2d_total}
P_{A^2}(a^2) = \sum_{i=1}^N p_i P_{A^2}^{i}(a^2) \,,
\end{equation}

\subsection{4D symbol mapping}
Since four consecutive amplitudes from a single shaped amplitude sequence are used to produce a 4D-amplitude vector (all components of 4D-amplitude vector are not independent), the 4D-quadrant PMF resulting from the $i^\mathrm{th}$ composition is given by
\begin{equation}\label{pmf4d_composition}
P_{A^4}^i(a^4) =
\bigg[\frac{c_{k_1}^i}{L} \cdot \frac{c_{k_2}^i-\delta_{1}}{L-1} \cdot \frac{c_{k_3}^i-\delta_{2}}{L-2} \cdot \frac{c_{k_4}^i-\delta_{3}}{L-3} \bigg]^{+}\,,  
\end{equation}
where $\delta_l$ is
\begin{equation}\label{pmf4d_correction}
\delta_{l} = \sum_{m=1}^{l} \textrm{I}_{k_{l+1} = k_m}\,,
\end{equation}
where $\textrm{I}_{k_{l+1} = k_m}$ is the indicator function which equals 1 when the condition ${k_{l+1} = k_m}$ is true and equals 0 otherwise. 

The total 4D-PMF is
\begin{equation}\label{pmf4d_total}
P_{A^4}(a^4) = \sum_{i=1}^{N}p_i P_{A^4}^i(a^4)\,.
\end{equation}
We note that the 4D-PMF can not be decomposed into the product of lower-dimensional PMFs.  

\subsection{Example}
Consider a shaper with a single composition $C = \{6,5,3,2\}$ of length $L=16$. The probability of 4D-symbol $x_1=[+7,+7,+7,+7]$ resulting from composition $C$ using 1D, 2D and 4D symbol mapping will be
\begin{align*}
P_{X}^{\mathrm{1D}}(x_1) &= \frac{1}{16} \cdot \frac{2}{16} \cdot \frac{2}{16} \cdot \frac{2}{16} \cdot \frac{2}{16} = 0.000015\,, \\[5pt]
P_{X}^{\mathrm{2D}}(x_1) &= \frac{1}{16} \cdot \frac{2}{16} \cdot \frac{1}{15} \cdot \frac{2}{16} \cdot \frac{1}{15} = 0.000004\,, \\[5pt]
P_{X}^{\mathrm{4D}}(x_1) &= \frac{1}{16} \cdot \frac{2}{16} \cdot \frac{1}{15} \cdot \frac{0}{14} \cdot \frac{0}{13} = 0\,.
\end{align*}
For $x_2=[+1,+3,+5,+7]$ the probabilities will be
\begin{align*}
P_{X}^{\mathrm{1D}}(x_2) &= \frac{1}{16} \cdot \frac{6}{16} \cdot \frac{5}{16} \cdot \frac{3}{16} \cdot \frac{2}{16} = 0.000172\,, \\[5pt]
P_{X}^{\mathrm{2D}}(x_2) &= \frac{1}{16} \cdot \frac{6}{16} \cdot \frac{5}{15} \cdot \frac{3}{16} \cdot \frac{2}{15} = 0.000195\,, \\[5pt]
P_{X}^{\mathrm{4D}}(x_2) &= \frac{1}{16} \cdot \frac{6}{16} \cdot \frac{5}{15} \cdot \frac{3}{14} \cdot \frac{2}{13} = 0.000258\,.
\end{align*}

We further note that higher-dimensional mapping reduces the probability of having equal amplitudes and increases the probability of having unequal amplitudes in simultaneous quadratures of a 4D-symbol. This leads to less probable high peak power values in the case of higher-dimensional mapping while maintaining the same power efficiency. 

\section{Experimental Data Fitting}
\label{SecondAppendix}
Data fitting based on the GN-model \cite{GN_model} was performed for each set of experimental points, which measure effective SNR and AIR at a specified optical launch power.   

\subsection{Effective SNR Fitting}
Effective SNR is the combination of ASE, nonlinear and transceiver SNR terms. It can be expressed as 
\begin{equation}
\label{snr_sum}
\begin{split}
\frac{1}{\rm{SNR}_{eff}} & = \frac{1}{\rm{SNR}_{ASE}} + \frac{1}{\rm{SNR}_{NLin}} + \frac{1}{\rm{SNR}_{Tr}}\\[5pt] 
& = \frac{a}{P} + \frac{b \cdot P^3}{P} + c\,,
\end{split}
\end{equation}
where ${\rm{SNR}_{ASE}}$ is a linear SNR term due to ASE in EDFAs, ${\rm{SNR}_{NLin}}$ is a nonlinear SNR term due to nonlinear fiber Kerr nonlinearity (nonlinear noise power is assumed to have cubic dependence on power \cite{GN_model}) and ${\rm{SNR}_{Tr}}$, which is the constant SNR term due to transceiver electrical noise and component imperfections; $P$ is the launch power; $a$, $b$, $c$ are the fitting parameters.       
Effective SNR can be rewritten as follows:
\begin{equation}
{\rm{SNR}_{eff}} =  \frac{P}{a + c \cdot P+b \cdot P^3} \,. 
\label{snr_fit}
\end{equation}

Initial estimation of fitting parameters is performed as follows: $c$ is estimated according to (\ref{c_guess}) using BtB measurements, $a$ is estimated using launch power sweep measurements neglecting the nonlinear term at low launch power values (\ref{a_guess}), and finally $b$ is estimated using launch power sweep measurements at high launch power values (\ref{b_guess}). 
\begin{align}
\begin{split}\label{c_guess}
c &= \frac{1}{\rm{SNR}_{eff}^{BtB}} - \frac{1}{\rm{SNR}_{ASE}^{BtB}}\\[5pt]
    &= \frac{1}{\rm{SNR}_{eff}^{BtB}} - \frac{BW}{\rm{OSNR} \cdot \textrm{12.5 GHz}}\,,
\end{split}\\[5pt]
\label{a_guess}
a &= \frac{P}{\rm{SNR}_{eff}^{Lin}} - c \cdot P\,, \\[5pt]
\label{b_guess}
b &= \frac{1}{{\rm{SNR}_{eff}^{NLin}} \cdot P^2} - \frac{a}{P^3} - \frac{c}{P^2}\,,
\end{align}
where $BW$ is the signal bandwidth. 

Refinement of the estimates of fitting parameters may be carried out, for example, using the least squares method \cite{datafitting}.
\subsection{AIR Fitting}
For AIR fitting we use a linear mapping with regard to SNR in a logarithmic scale  
\begin{equation}\label{air_fit}
{\rm{AIR}} = k \cdot \log_{10} ({\rm{SNR}_{eff}})\,,
\end{equation}
where $k$ is the fitting parameter. For the range of SNR values considered in this work, linear mapping between AIR and SNR (in logarithmic scale) shows very good agreement. However, we note that for very low or very high SNR values (when AIR saturates) this approach is unsuitable.    

\section*{Acknowledgment}

The experimental work carried out at Aston University was supported by UK EPSRC grant EP/M009092/1. We thank Lumentum UK for loan of the CFP2-ACO and studentship support of Pavel Skvortcov, and Socionext for loan of the DAC/ADC DKs used in this work.

\ifCLASSOPTIONcaptionsoff
  \newpage
\fi



\bibliographystyle{IEEEtran}
\bibliography{IEEEabrv,JSTQE_HCSS}
%



\end{document}